\documentclass[journal, a4paper, twoside]{IEEEtran}
\IEEEoverridecommandlockouts
\usepackage[utf8]{inputenc} 
\usepackage[T1]{fontenc}
\usepackage{url}
\usepackage{ifthen}
\usepackage[noadjust]{cite}
\usepackage[cmex10]{amsmath} 
\usepackage{amssymb}
\usepackage{graphicx}
\usepackage{epstopdf}
\usepackage{epsfig}
\usepackage{multirow}
\usepackage{amsthm}
\usepackage{comment}
\usepackage{caption}
\usepackage{enumitem}
\usepackage{amsmath}
\usepackage{color}
\usepackage{bbm}
\usepackage{algorithm}
\usepackage{algpseudocode}
\usepackage{csquotes}
\usepackage{subfiles}
\usepackage{caption, multirow, makecell}
\usepackage{array}
\usepackage{caption}
\usepackage{nicematrix}
\usepackage{cuted}
\usepackage{subcaption}
\usepackage{mathtools}
\usepackage{blkarray}

\DeclarePairedDelimiter\floor{\lfloor}{\rfloor}

\newtheorem{thm}{Theorem}
\newtheorem{lem}{Lemma}

\newtheorem{defn}{Definition}

\newtheorem{exmp}{Example}

\newtheorem{rem}{Remark}

\allowdisplaybreaks

\def\NoNumber#1{{\def\alglinenumber##1{}\State #1}\addtocounter{ALG@line}{-1}}

\def\BibTeX{{\rm B\kern-.05em{\sc i\kern-.025em b}\kern-.08em
    T\kern-.1667em\lower.7ex\hbox{E}\kern-.125emX}}

\begin{document}
\title{Multi-Antenna Coded Caching for Multi-Access Networks with Cyclic Wrap-Around }
\author{Elizabath Peter, \IEEEmembership{Member, IEEE,} K. K. Krishnan Namboodiri, 
	\IEEEmembership{Member, IEEE,}\\ and B. Sundar Rajan, \IEEEmembership{Life Fellow, IEEE} 
       \thanks{This work was supported partly by the Science and Engineering Research Board (SERB) of Department of Science and Technology (DST), Government of India, through J.C Bose National Fellowship to Prof. B. Sundar Rajan, and by the Ministry of Human Resource Development (MHRD), Government of India through Prime Minister's Research Fellowship to Elizabath Peter and K. K. Krishnan Namboodiri. A part of the content of this manuscript appeared in \emph{Proc. IEEE WCNC,}  2024, doi: 10.1109/WCNC57260.2024.10570537 \cite{PNR}.
       
       
      Elizabath Peter, K. K. Krishnan Namboodiri, and B. Sundar Rajan are with the Department of Electrical Communication Engineering, Indian Institute of Science, Bengaluru 560012, India (e-mail: elizabathp@iisc.ac.in, krishnank@iisc.ac.in, bsrajan@iisc.ac.in). 
}}
\maketitle
\begin{abstract}
This work explores a multiple transmit antenna setting in a multi-access coded caching (MACC) network where each user accesses more than one cache. A MACC network has $K$ users and $K$ caches, and each user has access to $r < K$ consecutive caches in a cyclic wrap-around manner. There are $L$ antennas at the server, and each cache has a normalized size $M/N \leq 1$. The cyclic wrap-around MACC network with a single antenna at the server has been well-investigated, and several coded caching schemes and improved lower bounds on the performance are derived for the same. However, this MACC network has not yet been studied under multi-antenna settings in the coded caching literature. We study the multi-antenna MACC problem and propose a solution for the same by constructing a pair of arrays called caching and delivery arrays. We present four constructions of caching and delivery arrays for different scenarios and obtain corresponding multi-antenna MACC schemes. Three of the above schemes achieve optimal performance under uncoded placement and one-shot delivery. The optimality is shown by matching the performance of the multi-antenna MACC scheme to the optimal performance in a dedicated cache network having $K$ users and  normalized cache size  $rM/N$. Further, as a special case, one of the proposed schemes subsumes an existing optimal MACC scheme for the single-antenna setting.
\end{abstract}

\section{Introduction}
Caching has been considered an efficient solution to tackle the needs of emerging wireless applications. Caching alleviates the high temporal variability of network traffic by prefetching contents into the caches distributed across the network during off-peak hours. When users simultaneously request files during peak times, these prefetched contents are then used to deliver a part of the requested files, thus reducing the amount of data that needs to be transmitted over the link and the total time required to serve the users. In conventional caching techniques, the server sends the remaining data in uncoded form. In \cite{MaN}, the authors illustrated that multicasting opportunities could be created by employing coded transmissions in the delivery phase. Thus, coded transmissions resulted in a further reduced delivery time compared to the one required in the case of uncoded caching techniques. The network model studied in \cite{MaN} is a dedicated cache network with a single-antenna server connected to a set of single-antenna users, each with its own cache, over an error-free shared link. For this network, a coded caching scheme was proposed in \cite{MaN}, which was later  shown to be optimal under uncoded placement and distinct user demands \cite{YMA,WTP}. Several follow-up works came since \cite{MaN} studying the coded caching approach in different settings \cite{HKD,YCT,STS,MKR,BEl,NaR,NaR2}.

In this work, our interest is in the well-explored multi-access coded caching (MACC) network \cite{HKD,NaR,ReK1,SPE,ReK2} having an identical number of users and caches, and each user is accessing more than one cache in a consecutive and cyclic wrap-around manner. Recently, different types of MACC networks were proposed that were mainly driven by designs \cite{KMR} and combinatorial topology \cite{MKR,BEl,NaR2}. The cyclic wrap-around MACC network studied in the literature consists of a single-antenna server connected to a set of $K$ users through a shared link. The server has $N$ files and there are $K$ caches in the network, each cache is having a size of $M \leq N$ files. Each user has access to $r$ consecutive caches in a cyclic wrap-around manner. This MACC network model was inspired by the widely studied ring networks, such as the circular Wyner model for interference networks with limited interference from neighbours \cite{WTS,SSPS}. Even though cyclic wrap-around MACC networks were studied extensively; there are only two schemes that achieve optimal performance \cite{SPE,NaR}. The scheme in \cite{NaR} holds only for $M \leq \frac{N-(K-r)}{K}$ and is optimal when $N \leq K$. The other scheme in \cite{SPE} achieves optimal performance under uncoded placement when $r=\frac{K-1}{KM/N}$. In all the previous works on cyclic wrap-around MACC networks, the server is assumed to have only a single antenna. The study on multiple transmit antenna settings in other network models have revealed that coded caching gain and multiplexing gain can be combined to reduce the delivery time achieved in the single-antenna setting \cite{NPR, YWCQC, LaE}. Therefore, it is natural to study cyclic wrap-around MACC networks with multiple transmit antennas and investigate whether a similar reduction is possible in this case. This work is the first to explore multi-antenna setting in cyclic wrap-around MACC networks and the technical contributions are summarized below.
\begin{itemize}
	\item We first introduce two arrays called caching array and delivery array that can together describe a multi-antenna coded caching scheme for a cyclic wrap-around MACC network. The caching array determines the content placement in the caches, and the delivery array indicates the contents known to the users and the delivery policy (the set of users and the subfiles served in each transmission). We show that there exists a multi-antenna MACC scheme corresponding to every pair of caching and delivery arrays (Section~\ref{subsec:cdarrays}). All the multi-antenna MACC schemes presented in this work are obtained by constructing appropriate caching and delivery array pairs.
	\item Four constructions of caching and delivery arrays are proposed. We define $t \triangleq KM/N \in \mathbb{Z}$.
	
	 \begin{enumerate}[label=(\roman*)]
    	\item  The first construction gives a caching array and a delivery array for any $K$, $r$, $t$, and $L$ satisfying the condition $K \geq r(t+L)$. The multi-antenna MACC scheme resulting from this set of caching and delivery arrays achieves a normalized delivery time $\frac{K-rt}{t+L}$ 
    	(Normalized delivery time is defined later in Section~\ref{sec:sysmodel}). (Section~\ref{subsec:const1}, Theorem~\ref{thm_general}).
    
    	\item The second construction is for the following case: $K =rt +L$ and $\gcd(K,t)=1$ (Section~\ref{subsec:const2_3}, Theorem~\ref{thm1}, case (a)). The multi-antenna MACC scheme obtained from this construction achieves the normalized delivery time $\frac{K-rt}{rt+L}$, which is optimal under uncoded placement and one-shot delivery. Even when $\gcd(K,t) \neq 1$, the above performance is achievable in certain scenarios (Section~\ref{subsec:non_unity gcd}).
    	
    	\item The third construction considers the case $K = mrt + (m-1)L$, where $L \geq rt$, $\gcd(K,t)=1$, and $m$ is an integer greater than one (Section~\ref{subsec:const2_3}, Theorem~\ref{thm1}, case (b)). The multi-antenna MACC scheme resulting from this construction also achieves the optimal normalized delivery time $\frac{K-rt}{rt+L}$. The scheme is also applicable when $\gcd(K,t,L) \neq 1$.
    	
    	The second and the third constructions are obtained by modifying the constructions of extended placement delivery arrays (EPDAs) given in \cite{NPR}.
    	
    	\item The fourth construction presents a systematic procedure to obtain a caching array and a delivery array from any given EPDA. The construction results in a multi-antenna scheme for an MACC network satisfying  the following conditions: $r$ divides $K$ and $r$ divides $L$  (Section \ref{subsec:const4}). The obtained scheme exhibits optimal performance under uncoded placement and one-shot delivery. Note that the conditions  $r$ divides $K$ and $r$ divides $L$  are independent of $t$. Thus, for an MACC network with $K$ and $L$ being multiples of $r$, the scheme characterizes optimal normalized delivery time versus memory tradeoff under the constraints of uncoded placement and one-shot delivery. (Section \ref{subsec:const4}, Theorem \ref{thm3}).
    	

    \end{enumerate}

    \item The scheme proposed in case (ii) generalizes the single-antenna MACC scheme in \cite{SPE} to an $L-$antenna setting (Section~\ref{subsec:const2_3}, Remark~\ref{remark1}). Note that the scheme in \cite{SPE} is optimal under uncoded placement for the single-antenna setting with $r=(K-1)/t$.
    
    \item When $L=1$, the MACC scheme obtained from Theorem~\ref{thm_general} achieves the normalized delivery time $\frac{K-rt}{t+1}$ with a lower subpacketization level (number of subfiles to which a file is divided) requirement than the MACC scheme in \cite{CWLZC}. 

\end{itemize}
\textit{Notations:}
The set $\{1,2,\ldots,n\}$ is denoted by $[n]$, and the set $\{m,m+1,\ldots,n\}$ is denoted by $[m,n]$. For any two integers $i$ and $K$, 
$$
\begin{small}
 \langle i \rangle _K  = \begin{cases} i\textrm{ }(mod\textrm{ } K)  & \textrm{ if }  i \textrm{ }(mod\textrm{ } K) \neq 0,\\
 K & \textrm{ if }  i \textrm{ }(mod\textrm{ } K) = 0.
 \end{cases}
 \end{small}
$$
The bold uppercase and lowercase letters denote matrices and vectors, respectively. The sets are denoted by calligraphic letters. For any set $\mathcal{S}$, $|\mathcal{S}|$ denotes its cardinality. For an $m \times n$ matrix $\mathbf{A}=[\mathbf{a}_1,\mathbf{a}_2,\ldots,\mathbf{a}_n]$, the sub-matrix $[\mathbf{a}_1,\mathbf{a}_2,\ldots,\mathbf{a}_{n^{\prime}}]$, where $n^{\prime} <n$, is denoted by $\mathbf{A}_{[1:n^{\prime}]}$. For two vectors $\mathbf{u}$ and $\mathbf{v}$, $\mathbf{u} \perp \mathbf{v}$ means that $\mathbf{u}^\mathrm{T}\mathbf{v}=0$. For two positive integers $a$ and $b$, $a$ divides $b$ is denoted by $a \mid b$.
\begin{figure}[t!]
	\begin{center}
		\captionsetup{justification=centering}
		\includegraphics[width=0.7\columnwidth]{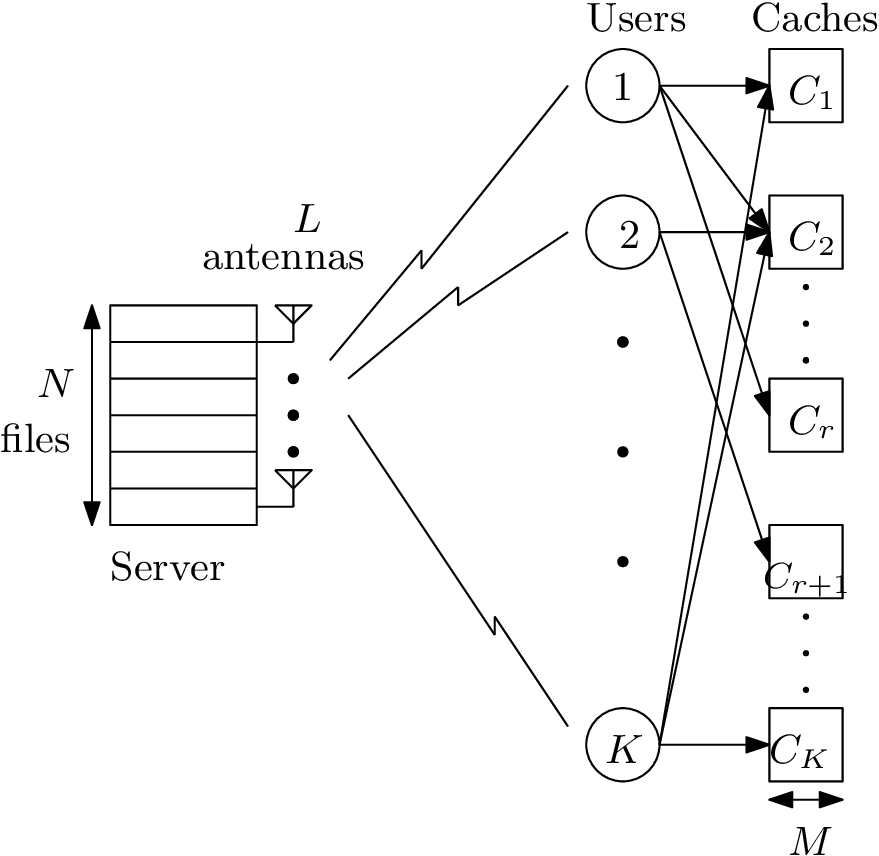}
		\caption{Cyclic wrap-around MACC network with multiple transmit antennas.}
		\label{fig:setting}
	\end{center}
\end{figure}

\section{System and Channel Model}
\label{sec:sysmodel}
The multi-antenna MACC network consists of a server with $L$ transmit antennas having access to a library of $N$ files $W^{[N]}=\{W^1,W^2,\ldots,W^N\}$, each of size $B$ bits, connected to $K$ users through a wireless broadcast link. There are $K$ caches distributed in the network, each of size $M$ files, where $0\leq M \leq N$. Each user has access to $r$ consecutive caches in a cyclic wrap-around manner. The $i^{th}$ cache is denoted by $C_{i}$, where $i \in [K]$. The set of caches accessible to user $k \in [K]$ is denoted by $\mathcal{A}_k$ and is given as $\mathcal{A}_k=\{C_i: i \in [k,\langle k+r-1 \rangle_{K}]\}$. Thus, $|\mathcal{A}_k|=r$. An illustration of the network model is given in Fig.~\ref{fig:setting}. The communication over this network happens in two phases called placement and delivery phases. In the placement phase, the caches are filled with the file contents subject to the memory constraint. The contents stored in the $i^{th}$ cache are denoted by $\mathcal{Z}_{C_i}$, and $\mathcal{Z}_{C_i}$ is a function of $W^{[N]}$ such that $|\mathcal{Z}_{C_i}| \leq M$. Therefore, the file contents available to user $k$, denoted by $Z_{k}$, are: $Z_k=\cup_{i \in \mathcal{A}_k}\mathcal{Z}_{i}$, which implies $|Z_k| \leq rM$ files.

The delivery phase starts when the users demand one of the $N$ files from the server during peak times. Let the demand of the $k^{th}$ user be denoted as $d_k$. Then, for a demand vector $\mathbf{d}=(d_1,d_2,\ldots,d_K)$, the server makes a set of transmissions as defined by the delivery policy. Let $\mathbf{x} \in \mathbb{C}^{L \times 1}$ be a transmitted message. Then, corresponding to $\mathbf{x}$, the received message at user $k$ is of the form:
\begin{equation*}
{y}_k = \mathbf{h}_k^{\mathrm{T}}\mathbf{x} + n_k,
\end{equation*}
where $\mathbf{h}_k=[h_{k,1}, h_{k,2}, \ldots, h_{k,L}]^{\mathrm{T}}$ represents a random vector of channel gains from $L$ antennas to user $k$, and $n_k$ is the additive white gaussian noise with unit variance observed at user $k$. The transmitted message $\mathbf{x}$ follows the power constraint $\mathbb{E}[||\mathbf{x}||^2] \leq P$. The channel gains are known to all the users and the server, and it is assumed to be constant during each transmission. Each user can retrieve the demanded file using the received messages and the side-information. Let $\widehat{W}^{d_k}$ be the file decoded by user $k$ at the end of the delivery phase. Then, for the given placement and delivery policy, the worst-case probability of error, $P_e$, is defined as 
\begin{equation*}
 P_e=\max_{\mathbf{d} \in [N]^K}\max_{k \in [K]}\mathbb{P}(W^{d_k} \neq \widehat{W}^{d_k}).
\end{equation*}
A coded caching scheme is said to be achievable if $P_e \rightarrow 0$ when $B \rightarrow \infty$ for all possible channel realizations. We consider high  signal-to-noise ratio (SNR) regime and perform a degrees of freedom (defined later) analysis as  done in \cite{NPR,YWCQC, LaE,STS2,CTXL,MoB}. In this paper, we do not delve into the discussion of finite SNR performance of the schemes as done in \cite{TSKK,SCK}. The performance measure is normalized delivery time, $T_n$, an information-theoretic metric,  which is defined as the worst-case delivery time required to satisfy any possible user demand $\mathbf{d}$, normalized with respect to the time taken to transmit a file of size $B$ bits to a single user at a rate $\log P$ in the high SNR regime under no caching scenario \cite{STS2}. i.e.,
 \begin{equation}
  T_n = \lim_{P \rightarrow \infty}\lim_{B \rightarrow \infty}\frac{\max_{\mathbf{d} \in [N]^K} T(\mathbf{d})}{B/\log P}.
 \end{equation}
Thus, the normalized delivery time represents the time required for all the users to decode its file correctly under sufficiently large file size and high SNR. The optimal normalized delivery time, $T^{*}_n$, is then defined as
\begin{equation}
  T^{*}_n = \inf \{T_n: T_n \textrm{ is achievable}\}.
\end{equation}
The degrees of freedom (DoF) achieved by a multi-access coded caching scheme is defined as $K(1-rM/N)/T_n$. The objective of any coded caching scheme is to minimize $T_n$, or, in other words, maximize the DoF. When we minimize $T_n$, the spectral efficiency of the system also gets improved.
\section{Preliminaries}
\label{prelims}
In this section, we give a brief review on extended placement delivery array (EPDA) introduced in \cite{NPR}, \cite{YWCQC} for the design of multi-antenna coded caching schemes. Corresponding to an EPDA, there exists a multi-antenna coded caching scheme for the dedicated cache network. 
\begin{defn}[Extended Placement Delivery Array \cite{NPR,YWCQC}]
	\label{defn:ata}
	Let $K,L (\leq K), F,Z, S$ be positive integers. An array $\mathbf{A}=[a_{j,k}]$, $j\in [F]$, $k\in [K]$ consisting of the symbol $\star$ and positive integers in $[S]$ is called a $(K,L,F,Z,S)$ extended placement delivery array (EPDA) if it satisfies the following conditions:\\
	C1. The symbol $\star$ appears $Z$ times in each column.\\
	C2. Every integer in the set $[S]$ occurs at least once in $\mathbf{A}$.\\ 
	C3. No integer appears more than once in any column.\\
	C4. Consider the sub-array $\mathbf{A}^{(s)}$ of $\mathbf{A}$ obtained by deleting all the rows and columns of $\mathbf{A}$ that do not contain the integer $s$. Then for any $s\in [S]$, no row of $\mathbf{A}^{(s)}$ contains more than $L$ integers.
\end{defn}
If each integer in the EPDA appears exactly $g-$
times, then the EPDA is said to be $g-$regular.
 \begin{thm}[\cite{NPR}]
	\label{thm:ata}
	Corresponding to any $(K,L,F,Z,S)$ EPDA, there exists a multi-antenna coded caching scheme for the dedicated cache network with $K$ users, $L$ transmit antennas, and ${M}/{N}={Z}/{F}$. Furthermore, the server can meet any user demand $\mathbf{d}$ with an NDT $T_n = {S}/{F}$.
\end{thm}
The EPDAs are introduced to obtain multi-antenna coded caching schemes with lower subpacketization requirements. 

\section{Main Results}
\label{sec:main_result}
In this section, we present multi-antenna coded caching schemes for the cyclic wrap-around MACC network discussed in Section~\ref{sec:sysmodel}. To describe the schemes, we first define two arrays called \textit{caching array} and \textit{delivery array}. 
\subsection{Caching and Delivery Arrays}
\label{subsec:cdarrays}
\begin{defn}[Caching array]
 For positive integers $K$, $F$, $Z$, and $r$, an array $\mathbf{C}=[c_{j,k}]$, $j \in [F], k \in [K]$, is called a $(K,F,Z,r)$ caching array if the following conditions are satisfied.\\
 \textit{B1.} Each column in $\mathbf{C}$ should contain $Z$ symbol  $\star$'s.\\
 \textit{B2.} In $\mathbf{C}$, the support of $\star$'s in any two columns $\mathbf{c}_{k_1}$ and $\mathbf{c}_{k_2}$, satisfying $\langle k_1-k_2 \rangle_K<r$ or $\langle k_2-k_1\rangle_K<r$, should be disjoint, i.e., if $\mathcal{R}_{k_1}\triangleq \{j: c_{j,k_1}=\star, \textrm{\hspace{0.1cm}} \forall j \in [F]\}$ and $\mathcal{R}_{k_2}\triangleq \{j: c_{j,k_2}=\star, \textrm{\hspace{0.1cm}} \forall j \in [F]\}$, then $|\mathcal{R}_{k_1} \cap \mathcal{R}_{k_2}|=0$ for a pair of columns satisfying the above property.
\end{defn}
Thus, the $(K,F,Z,r)$ caching array $\mathbf{C}$ is an $F \times K$ array consisting of only $\star$'s and nulls, and satisfies both the conditions $B1$ and $B2$.

\begin{defn}[Delivery array]
	\label{def:delarray}
	Given a $(K,F,Z,r)$ caching array $\mathbf{C}$, and positive integers $S$, $L$, an array $\mathbf{D}=[d_{j,k}]$, $j \in [F]$, $k \in [K]$, composed of a symbol `$\star$' and $S$ positive integers from $[S]$, is called a $(\mathbf{C},S,L)$ delivery array if it satisfies the below conditions:\\
	\textit{D1.} The position of $\star$'s in each column of $\mathbf{D}$ is defined by the caching array $\mathbf{C}$. For a column $\mathbf{d}_k$, 
	\begin{equation*}
	   d_{j,k} = \star, \textrm{\hspace{0.1cm}}\forall j \in \cup_{l \in [k,\langle k+r-1\rangle_K]}\mathcal{R}_l,
	\end{equation*}
	where $\mathcal{R}_l \triangleq \{j: c_{j,l}=\star, \forall j \in [F]\}$. \\
	\textit{D2. } Every integer in the set $[S]$ occurs at least once in $\mathbf{D}$, and not more than once in any column.\\
	\textit{D3. } Any row in the sub-array $\mathbf{D}^{(s)}$, obtained after eliminating the rows and columns not containing the integer $s$, has at most $L$ integers.
\end{defn}

To further illustrate, an example of a $(7,7,3,2)$ caching array $\mathbf{C}$ and a $(\mathbf{C},3,3)$  delivery array $\mathbf{D}$ are given in \eqref{eq: cacexmp} and \eqref{eq:delexmp}, respectively.
\begin{equation}
	\begin{footnotesize}{
	\mathbf{C}=\begin{blockarray}{ccccccc}
		1 & 2 & 3 & 4 & 5 & 6 & 7 \\
		\begin{block}{[ccccccc]}
			\star & & & \star & & &  \\
			\star & & & & \star & &  \\ 
			& \star & & & \star & &  \\ 
			& \star & & & & \star &  \\ 
			& & \star  & & & \star &  \\ 
			& & \star  & & & & \star   \\ 
			& & & \star  & & & \star   \\ 
		\end{block}
	\end{blockarray} }\end{footnotesize}
\label{eq: cacexmp}
\end{equation}
\begin{equation}
	\begin{footnotesize}{
	\mathbf{D}=\begin{blockarray}{ccccccc}
		1 & 2 & 3 & 4 & 5 & 6 & 7 \\
		\begin{block}{[ccccccc]}
			\star & 2 & \star & \star & 3 & 1 &  \star \\
			\star & 3 & 1 & \star & \star & 2 &  \star \\ 
			\star & \star & 2 & \star & \star & 3 & 1  \\ 
			\star & \star & 3 & 1 & \star & \star &  2 \\ 
			1 & \star & \star  & 2 & \star & \star &  3\\ 
			2 & \star & \star  & 3 & 1 & \star & \star   \\ 
			3 & 1 & \star & \star  & 2 & \star & \star   \\ 
		\end{block}
	\end{blockarray} } \end{footnotesize}
 \label{eq:delexmp}
\end{equation}
The position of $\star$'s in $\mathbf{D}$ is determined by the caching array $\mathbf{C}$ that we begin with. The number of $\star$'s present in each column of $\mathbf{D}$ is $rZ$ by virtue of condition $B2$ of the caching array. The integers are filled according to conditions $D2$ and $D3$. Note that the delivery array $\mathbf{D}$ in Definition~\ref{def:delarray} qualifies to be a $(K,L,F,rZ,S)$ EPDA. Based on the insights obtained from the EPDA constructions in \cite{NPR}, we construct caching and delivery arrays to derive multi-antenna coded caching schemes for cyclic wrap-around MACC networks. The following lemma gives the relation between multi-antenna MACC scheme and the arrays $\mathbf{C}$ and $\mathbf{D}$.
 
 \begin{lem}
 From a $(K,F,Z,r)$ caching array $\mathbf{C}$ and a $(\mathbf{C},S,L)$ delivery array $\mathbf{D}$, a multi-antenna coded caching scheme can be obtained for a cyclic wrap-around MACC network having $K$ users, each one accessing $r$ consecutive caches, $L$ transmit antennas at the server, and $M/N=Z/F$. The resulting scheme has the following performance measures: normalized delivery time $T_n=S/F$ and subpacketization level is $F$. 
 \label{ref:lem}
 \end{lem}
\begin{IEEEproof}
The multi-antenna coded caching scheme resulting from a $(K,F,Z,r)$ caching array $\mathbf{C}$ and a $(\mathbf{C},S,L)$ delivery array $\mathbf{D}$ is described below.

 \textit{Placement phase}: The placement is performed using the caching array $\mathbf{C}$. The columns of $\mathbf{C}$ represent the caches. Each file $W^n$, $\forall n \in [N]$, is divided into $F$ subfiles of equal size: $W^n = \{W^n_1,W^n_2,\ldots,W^n_F\}$. Each row, $j \in [F]$, in $\mathbf{C}$ represents the subfiles $\{W^n_j, \forall n \in [N]\}$. 
 The contents stored in the $i^{th}$ cache are given as:
 \begin{equation*}
  \mathcal{Z}_{C_i} = \{W^n_j, \forall n \in [N]: c_{j,i}=\star, j \in [F]\}.
 \end{equation*}
 The total number of subfiles stored in each cache is $NZ$, and each subfile has a normalized size $1/F$. Thus, $M/N=Z/F$. 
 
\textit{Delivery phase:} The columns of $\mathbf{D}$ represent the users, and the $\star$'s in each column represent the subfiles known to each user. Each user knows $rZ$ subfiles of a file. Thus, the contents available to the $k^{th}$ user are ${Z}_k = \{W^n_j, \forall n \in [N]: d_{j,k}=\star, j \in [F]\}$. Condition B2 ensures that there is no redundancy in the cache contents accessible to a user. Thus, a user gets full local caching gain by the above placement.

On receiving a demand vector $\mathbf{d}=(d_1,d_2,\ldots,d_K)$, the server sends messages  corresponding to every distinct integer present in $\mathbf{D}$. Thus, there are $S$ transmissions, each of size $B/F$ bits. Therefore, we get $T_n = S/F$. 
 
 Next, we show the correctness of the above delivery scheme. Consider an integer $s \in [S]$ present in $\mathbf{D}$. Let $l$ denote the count of $s$ in $\mathbf{D}$. Find the sub-array $\mathbf{D}^{(s)}$, and let $d_{j_1,k_1}=d_{j_2,k_2}=\cdots=d_{j_l,k_l}=s$. Then, the transmission corresponding to $s$ is given by  $\mathbf{V}^{(s)}[W^{d_{k_1}}_{j_1},W^{d_{k_2}}_{j_2},\ldots,W^{d_{k_l}}_{j_l}]^{\mathrm{T}}$, where $\mathbf{V}^{(s)}=[\mathbf{v}^{(s)}_{k_1}, \mathbf{v}^{(s)}_{k_2},\ldots, \mathbf{v}^{(s)}_{k_l}]$ is a precoding matrix of size $L \times l$. The vector $\mathbf{v}^{(s)}_{k_i} \in \mathbb{C}^{L \times 1}$, $i \in [l]$, is designed such that $\mathbf{v}_{k_i} \perp \mathbf{h}_{u}$, where $u \in \{k: d_{j_i,k} \neq \star, k \in \{k_1,\ldots,k_{i-1},k_{i+1},\ldots,k_l \}\}$. At user $k_i$, the received message ${y}_{k_i}^{(s)}$ corresponding to the above transmission is 
 \begin{equation*}
    {y}_{k_i}^{(s)}=\mathbf{h}_{k_i}^{\mathrm{T}}\mathbf{V}^{(s)}[W^{d_{k_1}}_{j_1},W^{d_{k_2}}_{j_2},\ldots,W^{d_{k_l}}_{j_l}]^{\mathrm{T}} + n_{k_i}.
 \end{equation*}
Due to the high SNR assumption, we neglect $n_{k_i}$ in the further analysis. Then, 
\begin{equation*}
\begin{aligned}
  {y}_{k_i}^{(s)}= \mathbf{h}_{k_i}^{\mathrm{T}}\mathbf{v}^{(s)}_{k_i}W^{d_{k_i}}_{j_i}&+\underbrace{\sum_{\substack{u=1,\\u \neq i,d_{j_u,k_i}\neq \star}}^{l}\mathbf{h}_{k_i}^{\mathrm{T}}\mathbf{v}^{(s)}_{k_u}W^{d_{k_u}}_{j_u}}_{=0} \notag \\
  & +\underbrace{\sum_{\substack{u=1,\\u \neq i,d_{j_u,k_i}= \star}}^{l}\mathbf{h}_{k_i}^{\mathrm{T}}\mathbf{v}^{(s)}_{k_u}W^{d_{k_u}}_{j_u}}_\text{Using $Z_{k_i}$, user $k_i$ can compute it}.
\end{aligned}
 \end{equation*}
\noindent The design of precoding vectors enables to null the subfiles that are not available in $Z_{k_i}$. Thus, user $k_i$ can decode its desired subfile $W^{d_{k_i}}_{j_i}$ after removing the terms involving the known subfiles. Likewise, each user is able to retrieve its desired subfiles from the transmissions. In fact, the delivery policy and the decodability follow from the EPDAs in \cite{NPR}.
\end{IEEEproof}

The multi-antenna coded caching scheme resulting from the caching and delivery arrays follows an uncoded placement and one-shot delivery. A delivery scheme is said to be one-shot if each user is able to decode its desired subfile in at most one channel use. Later in the work, we prove that the performance of some of the proposed schemes is optimal under uncoded placement and one-shot delivery. The optimality of the proposed schemes is shown by matching their performance to the optimal performance in the equivalent multi-antenna dedicated cache setting. The dedicated cache network equivalent to the cyclic wrap-around MACC network shown in Fig. \ref{fig:setting} has $K$ users, $N$ files, $L$ antennas, and a normalized cache size $rM/N$. The optimal normalized delivery time under uncoded placement and one-shot delivery for the above dedicated cache network is shown to be $\frac{K-rt}{rt+L}$, where $t = KM/N \in [0,\floor{\frac{K}{r}}]$ and $rt+L \leq K$ \cite{LBE}. If we denote the optimal performance under uncoded placement and one-shot delivery in the cyclic wrap-around MACC setting by $T^{*}_{n,u}$, then
\begin{equation}
T^{*}_{n,u} \geq \frac{K-rt}{rt+L}.
\label{eq:bound}
\end{equation}

\subsection*{Novel constructions of caching and delivery arrays}
\label{subsec:schemes}
We present four constructions of caching and delivery arrays, each one considering a different setting.
The main objective behind those constructions is to obtain caching and delivery arrays with lower $F$ and $S/F$ values.  Define $t \triangleq \frac{KM}{N} \in  [0,\floor{\frac{K}{r}}]$.

 \subsection{Construction I}
 \label{subsec:const1}
  Firstly, we present a construction of caching array and delivery array that applies for almost all values of $K$, $t$, $r$, and $L$. The resulting  multi-antenna MACC  scheme achieves full local caching gain and a DoF  $t+L$.  Thus, we have the following theorem.

\begin{thm}
	For a cyclic wrap-around MACC network with $L$ transmit antennas at the server, the following normalized delivery time 
	\begin{equation}
	   T_n = \frac{K-rt}{t+L}
	   \label{eq:general_rate}
	\end{equation}
	is achievable, when $K \geq r(t+L)$.
	\label{thm_general}
\end{thm}

\begin{IEEEproof}	
	The scheme that achieves the performance in \eqref{eq:general_rate} is obtained by constructing two arrays, a $(K,\binom{K-(r-1)(t+L)-1}{t+L-1}K,\binom{K-(r-1)(t+L)-1}{t+L-1}t,r)$ caching array $\mathbf{C}$ and a $(\mathbf{C},\frac{K(K-rt)}{t+L}\binom{K-(r-1)(t+L)-1}{t+L-1},L)$ delivery array $\mathbf{D}$. The construction holds when $K \geq r(t+L)$.
	
	The columns of caching and delivery arrays are indexed by $k \in [K]$. In this construction, the rows of caching and delivery arrays are indexed using two sets $\mathcal{J} \subseteq [K]$ and $\mathcal{W}_{\mathcal{J}}$, which are defined as:
	\begin{subequations}
	\begin{align}
	  & \mathcal{J}  \triangleq \{j_1,j_2,\ldots,j_{t+L}\}\textrm{ such that } j_1 < j_2 < \cdots <j_{t+L}, \notag \\
	   &  \langle j_i - j_l \rangle_K \geq r \textrm{ and }  \langle j_l - j_i \rangle_K \geq r, \forall i,l \in [t+L], \label{eq:set1}\\
	 &  \mathcal{W}_{\mathcal{J}}  \triangleq \{w_{\mathcal{J}}\subset \mathcal{J}: w_{\mathcal{J}}=\{j_i,j_{\langle i+1 \rangle_{t+L}},\ldots, j_{\langle i+t-1 \rangle_{t+L}}\},\notag \\ & \textrm{\hspace{5cm}} \forall i \in [t+L]\} \label{eq:set2}.
	\end{align}
   \end{subequations}
	Thus, the set $\mathcal{W}_{\mathcal{J}}$ contains  $t+L$ sets formed by $t$ consecutive elements of $\mathcal{J}$. There are $\frac{K}{t+L}\binom{K-(r-1)(t+L)-1}{t+L-1}$ choices for $\mathcal{J}$, and each $\mathcal{J}$ results in $t+L$ $(\mathcal{J},w_{\mathcal{J}})$ pairs, where $w_{\mathcal{J}} \in \mathcal{W}_{\mathcal{J}}$. Let $\mathcal{I}$ be the set containing all possible $(\mathcal{J},w_{\mathcal{J}})$ pairs satisfying \eqref{eq:set1} and \eqref{eq:set2}. The rows of caching and delivery arrays are indexed by $(\mathcal{J},w_{\mathcal{J}}) \in \mathcal{I}$. Then, $F=|\mathcal{I}|=K\binom{K-(r-1)(t+L)-1}{t+L-1}$.
	
	 The constructions of caching and delivery arrays are given in Algorithm~\ref{alg_const}. The $(K,K\binom{K-(r-1)(t+L)-1}{t+L-1},t\binom{K-(r-1)(t+L)-1}{t+L-1},r)$ caching array $\mathbf{C}=[c_{(\mathcal{J},w_{\mathcal{J}}),k}]$, where $(\mathcal{J},w_{\mathcal{J}}) \in \mathcal{I}$, $k \in [K]$, is constructed according to lines $1$--$8$ of Algorithm~\ref{alg_const}. In the $k^{th}$ column of $\mathbf{C}$, the symbol `$\star$' is present in those rows with $w_{\mathcal{J}} \ni k$. There are $\binom{K-(r-1)(t+L)-1}{t+L-1}$ number of sets $\mathcal{J}$ with $\mathcal{J} \ni k$. Each such $\mathcal{J}$ results in $t$ number of $(\mathcal{J},w_{\mathcal{J}})$ pairs with $w_{\mathcal{J}} \ni k$. Thus, $Z=t\binom{K-(r-1)(t+L)-1}{t+L-1}$. The array $\mathbf{C}$ constructed according to lines 1--8 of Algorithm~\ref{alg_const} is a $(K,K\binom{K-(r-1)(t+L)-1}{t+L-1},t\binom{K-(r-1)(t+L)-1}{t+L-1},r)$ caching array as it satisfies both the conditions $B1$ and $B2$. Condition $B2$ directly follows from the way in which the sets $\mathcal{J}$ are defined.

	Next, we construct a $(\mathbf{C},\frac{K(K-rt)}{t+L}\binom{K-(r-1)(t+L)-1}{t+L-1},L)$ delivery array $\mathbf{D}$. The $\star$'s in $\mathbf{D}$ are placed according to $\mathbf{C}$. Since a user $k \in [K]$ has access to $r$ consecutive caches, $d_{(\mathcal{J},w_{\mathcal{J}}),k}=\star$
    if  $c_{(\mathcal{J},w_{\mathcal{J}}),l}=\star$, where $l \in [k,\langle k+r-1 \rangle_K]$. The set $\mathcal{J}$ ensures that the position of $\star$'s in the adjacent $r$ columns of $\mathbf{C}$ are disjoint. Thus, there are $rZ$ $\star$'s in each column of $\mathbf{D}$, and each user enjoys a full local caching gain.
    The remaining $K(F-rZ)$ vacant cells of $\mathbf{D}$ need to be filled using integers according to lines 16--22 of Algorithm~\ref{alg_const}. We arrange $(t+L)-$sized sets $\mathcal{J}$ defined in 
    \eqref{eq:set1} according to their lexicographic order, and define a function $\Phi_{t+L}(\mathcal{J})$ to be the order minus one. Similarly, we define another function $\Psi(\mathcal{J},i)$ which gives the position of $i$ in the ordered set $[K] \backslash \mathcal{J}_{acc}$, where $\mathcal{J}_{acc} \triangleq \big\{\{j,\langle j+1 \rangle_K,\ldots,\langle j+r-1 \rangle_K, \forall j \in \mathcal{J}\} \big\}$. Note that $\mathcal{J}_{acc}$ consists of the sets of caches accessible to each user in $\mathcal{J}$. Thus, $\Psi(\mathcal{J},i) \in [K-r(t+L)]$. Now, consider a column $k \in [K]$ and all those row indices $(\mathcal{J},w_{\mathcal{J}})$ such that $d_{(\mathcal{J},w_{\mathcal{J}}),k}\neq \star$. The integers are filled in three different ways according to the following conditions: (i) $k \in \mathcal{J}$ and $k \notin w_{\mathcal{J}}$, (ii) $\langle k+m \rangle_K \in \mathcal{J}$ and $\langle k+m \rangle_K \notin w_{\mathcal{J}}$, where $m \in [r-1]$, and (iii) $\langle k+m \rangle_K \notin \mathcal{J}$, $\forall m \in [0,r-1]$. First, let us consider case (i). For a fixed $k$ and $\mathcal{J} \ni k$,
    there are $L$ number of $(\mathcal{J},w_{\mathcal{J}})$ row indices satisfying the case (i) criterion, and the integers at all those positions $d_{(\mathcal{J},w_{\mathcal{J}}),k}$ are defined by line 17. It is clear that no two integers in a column filled according to line $17$ are the same. Similarly, for a fixed $k$ and $\mathcal{J}$, there exists $L$ number of $(\mathcal{J},w_{\mathcal{J}})$ pairs satisfying case (ii) criterion, and the integers at those positions $d_{(\mathcal{J},w_{\mathcal{J}}),k}$ are assigned according to line 19. In case (iii), the integers are  assigned according to line 21. Note that in case (iii), we have $\{k,\langle k+1 \rangle_K, \ldots, \langle k+r-1 \rangle_K\} \cap \mathcal{J} = \emptyset$, and we define $\mathcal{J}_{int}=\mathcal{J} \cup \{k\}=\{b_1,\ldots,b_{t+L+1}\}$. Let $b_l=k$ and $b_p=j_{\langle i-1 \rangle_{t+L}}$, where $w_\mathcal{J} = \{j_i,j_{\langle i+1 \rangle_{t+L}},\ldots,j_{\langle i+t-1 \rangle_{t+L}}\}$, $l,p \in [t+L+1]$. If $\langle p+1 \rangle_{t+L+1} \neq l$, construct another set $\mathcal{J}^{\prime}_{int}$ by decrementing all the entries in $\mathcal{J}_{int}$ at positions  $\langle p+1 \rangle_{t+L+1},\ldots,\langle l-1 \rangle_{t+L+1}$ by $(r-1)$. Otherwise, $\mathcal{J}^{\prime}_{int}=\mathcal{J}_{int}$. Then, obtain a new set $\mathcal{J}_{new} \triangleq \mathcal{J}^{\prime}_{int} \backslash \{b_p\}$. This set $\mathcal{J}_{new}$ is used in line $21$ to identify the integer in $d_{(\mathcal{J},w_{\mathcal{J}}),k}$. The function $\Psi(\mathcal{J}_{new},j_{\langle i-1 \rangle_{t+L}})$ returns the position of $j_{\langle i-1 \rangle_{t+L}}$ in the ordered set $[K] \backslash \mathcal{J}_{acc}$, where $\mathcal{J}_{acc} = \big\{\{j,\langle j+1 \rangle_K,\ldots,\langle j+r-1 \rangle_K, \forall j \in \mathcal{J}_{new} \}\big\}$.
    It is evident that the set $\mathcal{J}_{acc}$ does not contain $j_{\langle i-1 \rangle_{t+L}}$, and there are $K-r(t+L)$ such integers that are not present in $\mathcal{J}_{acc}$. Thus, corresponding to every set $\mathcal{J}$, there are $K-r(t+L)+L+(r-1)L=K-rt$ distinct integers. Hence, the total number of integers in $\mathbf{D}$ is  obtained as $S=\frac{K(K-rt)}{t+L}\binom{K-(r-1)(t+L)-1}{t+L-1}$.
    
    	\begin{algorithm}[t!]
    	\renewcommand{\thealgorithm}{1}
    	\caption{Constructions of caching and delivery arrays in Theorem~\ref{thm_general}}
    	\label{alg_const}
    	\begin{algorithmic}[1]
    		\Procedure{Cachingarray}{$K,\mathcal{I}$}       
    		\State $\mathbf{C}=[c_{(\mathcal{J},w_{\mathcal{J}}),k}]$, $(\mathcal{J},w_{\mathcal{J}}) \in \mathcal{I}$, $k \in [K]$
    		\For{$k \in [K]$}
    		\For{$(\mathcal{J},w_{\mathcal{J}}) \in [\mathcal{I}]$}
    		\State $c_{(\mathcal{J},w_{\mathcal{J}}),k} = \star$, if $k \in w_{\mathcal{J}}$
    		\EndFor
    		\EndFor
    		\EndProcedure
    		
    		\Procedure{Deliveryarray}{$\mathbf{C},K,L,r,t,\mathcal{I}$}
    		\State $\mathbf{D}=[d_{(\mathcal{J},w_{\mathcal{J}}),k}]$, $(\mathcal{J},w_{\mathcal{J}}) \in \mathcal{I}$, $k \in [K]$
    		\State $F=|\mathcal{I}|$ and $S = {(K-rt)F}/{(t+L)}$ 
    		\For{$(\mathcal{J},w_{\mathcal{J}}) \in [\mathcal{I}]$}
    		\For {$k \in [K]$}
    		\If{$c_{(\mathcal{J},w_{\mathcal{J}}),l} = \star$,  $l \in [k,\langle k+r-1 \rangle_{K}]$}
    		\State{$d_{(\mathcal{J},w_{\mathcal{J}}),k} = \star$ }
    		\ElsIf {$k \in \mathcal{J}$}
    		\State{$d_{(\mathcal{J},w_{\mathcal{J}}),k}=(K-rt)\Phi_{t+L}(\mathcal{J})+$}
    		\NoNumber{\hspace{4cm}$\langle i -l \rangle_{t+L}$},
    		\NoNumber{$\mathcal{J}=\{j_1,\ldots,j_l=k,\ldots,j_{t+L}\}$ and}
    		\NoNumber{$w_{\mathcal{J}}=\{j_i,j_{\langle i+1 \rangle_{t+L}},\ldots,j_{\langle i+t-1 \rangle_{t+L}}\}$}
    		\ElsIf {$\langle k+m \rangle_K \in \mathcal{J}$, $m \in [r-1]$}
    		\State{$d_{(\mathcal{J},w_{\mathcal{J}}),k}=(K-rt)\Phi_{t+L}(\mathcal{J})+mL+$}
    		\NoNumber{\hspace{3.3cm}$\langle i -l \rangle_{t+L}$},
    		\NoNumber{$\mathcal{J}=\{j_1,\ldots,j_l=k+m,\ldots,j_{t+L}\}$ and}
    		\NoNumber{$w_{\mathcal{J}}=\{j_i,j_{\langle i+1 \rangle_{t+L}},\ldots,j_{\langle i+t-1 \rangle_{t+L}}\}$}
    		\Else
    		\State{$d_{(\mathcal{J},w_{\mathcal{J}}),k}=(K-rt)\Phi_{t+L}(\mathcal{J}_{new})+rL+$}
    		\NoNumber{\hspace{3.2cm}$\Psi(\mathcal{J}_{new}, j_{\langle i-1 \rangle_{t+L}})$},
    		\NoNumber{$\mathcal{J}_{int}=\mathcal{J} \cup \{k\}=\{b_1,\ldots,b_{t+L+1}\}$, } \NoNumber{\hspace{-0.5cm}where $b_l=k,b_p=j_{\langle i-1 \rangle_{t+L}}$, $l,p \in [t+L+1]$}
    		\NoNumber{\textbf{if} $\langle p+1 \rangle_{t+L+1} \neq l$ \textbf{then}}
    		
    		\NoNumber{ $\mathcal{J}^{\prime}_{int} \leftarrow $ Decrement the entries in $\mathcal{J}_{int}$}
    		\NoNumber{ present  at positions $\langle p+1 \rangle_{t+L+1},$}
    		\NoNumber{$ \langle p+2 \rangle_{t+L+1}, \ldots, \langle l-1 \rangle_{t+L+1} $ by $(r-1)$}
    		\NoNumber{\textbf{else}}
    		\NoNumber{ $\mathcal{J}^{\prime}_{int} = \mathcal{J}_{int}$}
    		\NoNumber{\textbf{end if}}
    		\NoNumber{$\mathcal{J}_{new} = \mathcal{J}^{\prime}_{int} \backslash \{{j_{\langle i-1 \rangle_{t+L}}}\}$}
    		\EndIf
    		
    		\EndFor
    		\EndFor    
    		\EndProcedure
    	\end{algorithmic}
    \end{algorithm}

     We now show that the array $\mathbf{D}$ constructed according to lines 9--25 satisfy all the three conditions of a delivery array. Lines 13--15 ensure that the placement of $\star$'s in $\mathbf{D}$ satisfies condition $D1$. To verify condition $D2$, let us first consider cases (i) and (ii). For a fixed $k$ and $\mathcal{J}$, the term $\langle i-l \rangle_{t+L}$ in lines 17 and 19 ensures that the same integer does not appear more than once in the $k^{th}$ column. Similarly, in case (iii), for a given $k \in [K]$, no two $(\mathcal{J},w_{\mathcal{J}})$ pairs result in same $\mathcal{J}_{new}$ and $\Psi(\mathcal{J}_{new},j_{\langle i-1 \rangle_{t+L}})$ together (line 21). Hence, an integer assigned according to line 21 does not appear more than once in a column. Thus, condition $D2$ is satisfied. Next, let us verify condition $D3$ case by case. Let $d_{(\mathcal{J},w_{\mathcal{J}}),k}$ be an integer $s \in [S]$ assigned according to line 17. The integer $s$ appears $t+L$ times in $\mathbf{D}$, specifically, in $t+L$ distinct rows and columns. Therefore, $\mathbf{D}^{(s)}$ is a $(t+L)\times(t+L)$ array formed by the rows and columns of $\mathbf{D}$ containing $s$. Then, the columns of $\mathbf{D}^{(s)}$ is constituted by the entries in $\mathcal{J}$, and therefore, each row of $\mathbf{D}^{(s)}$ contains $t$ $\star$'s. Thus, condition $D3$ is satisfied in case (i). Next, assume that $d_{(\mathcal{J},w_{\mathcal{J}}),k}=s$, where $s \in [S]$ is assigned according to line 19. Then, for some $m \in [r-1]$, let $j=\langle k + m \rangle_K \in \mathcal{J}$, the integer $s$ also appears $t+L$ times in $\mathbf{D}$. i.e., for every $j \in \mathcal{J}$, the integer $s$ appears in the $\langle j-m \rangle_K^{th}$ column of $\mathbf{D}$. Then, the sub-array $\mathbf{D}^{(s)}$ formed by the above columns and the rows containing $s$ is of size $(t+L) \times (t+L)$, and each row of $\mathbf{D}^{(s)}$ contains $t$ $\star$'s  as $d_{(\mathcal{J},w_{\mathcal{J}}),k}=\star$ when $c_{(\mathcal{J},w_{\mathcal{J}}),\langle k+m \rangle_K}=\star$, for $m \in [r-1]$ (from line 14). Hence, the number of integers in each row of $\mathbf{D}^{(s)}$ is $L$, thus satisfying $D3$. Now, it remains to verify $D3$ for integers assigned according to line 21. Consider a $(t+L)-$sized set $\mathcal{J}_{new}$, and a column index $u \in [K] \backslash \{\langle j + i \rangle_K: \forall j \in \mathcal{J}_{new},i \in [0,r-1]\}$. Now, form an ordered set $\mathcal{J}_{new} \cup \{u\}=\{a_1,a_2,\ldots,a_{t+L+1}\}$, and pick a $k \in \mathcal{J}_{new}$. Let $k=a_l$ and $u = a_p$, where $l,p \in [t+L+1]$. If $\langle p+1 \rangle_{t+L+1} \neq l$, increment all the entries in the set $\mathcal{J}_{new}\cup \{u\}$ present at positions $\langle p+1 \rangle_{t+L+1},\langle p+2 \rangle_{t+L+1},\ldots,\langle l-1 \rangle_{t+L+1}$ by $(r-1)$, and let the obtained set be called as $\mathcal{J}^{\prime}_{int}$. When $\langle p+1 \rangle_{t+L+1} = l$, we have $\mathcal{J}^{\prime}_{int}=\mathcal{J}_{new}\cup \{u\}$. For an integer $s \in [S]$ assigned according to case (iii), the row index $(\mathcal{J}^{(k)},w_{\mathcal{J}^{(k)}})$ for which $d_{(\mathcal{J}^{(k)},w_{\mathcal{J}^{(k)}}),k}=s$ is thus obtained as $\mathcal{J}^{(k)}=\mathcal{J}^{\prime}_{int} \backslash \{k\}=\{\alpha_1,\ldots,\alpha_q=u,\ldots,\alpha_{t+L}\}$ and  $w_{\mathcal{J}^{(k)}}=\{\alpha_{\langle q+1 \rangle_{t+L}},\alpha_{\langle q+2 \rangle_{t+L}},\ldots,\alpha_{\langle q+t \rangle_{t+L}}\}$. Notice that the sub-array $\mathbf{D}^{(s)}$ is of size $(t+L) \times (t+L)$, and the columns in $\mathbf{D}^{(s)}$ correspond to the entries in $\mathcal{J}_{new}$. When $d_{(\mathcal{J}^{(k)},w_{\mathcal{J}^{(k)}}),k}=s$, there exists $t$ number of $k^{\prime} \in \mathcal{J}_{new}$ such that $d_{(\mathcal{J}^{(k)},w_{\mathcal{J}^{(k)}}),k^{\prime}}=\star$ as  $w_{\mathcal{J}^{(k)}} \ni k^{\prime}$ or	$w_{\mathcal{J}^{(k)}} \ni \langle k^{\prime}+r-1 \rangle_{K} $. Therefore, $\mathbf{D}^{(s)}$ contains $t$ $\star$'s in each row indexed by $(\mathcal{J}^{(k)},w_{\mathcal{J}^{(k)}})$, where $k \in \mathcal{J}_{new}$. Thus, condition $D3$ is satisfied in case (iii) as well. 
     
     Using the above constructed $(K,K\binom{K-(r-1)(t+L)-1}{t+L-1},t\binom{K-(r-1)(t+L)-1}{t+L-1},r)$
     caching array $\mathbf{C}$ and $(\mathbf{C},\frac{K(K-rt)}{t+L}\binom{K-(r-1)(t+L)-1}{t+L-1},L)$ delivery array $\mathbf{D}$, we invoke Lemma~\ref{ref:lem} to obtain a multi-antenna scheme for the cyclic wrap-around MACC network satisfying $K \geq r(t+L)$ (the condition $K \geq r(t+L)$ arises from the binomial coefficient $\binom{K-(r-1)(t+L)-1}{t+L-1}$). The resulting multi-antenna coded caching scheme achieves the normalized delivery time $T_n=\frac{S}{F}=\frac{K-rt}{t+L}$, and has the subpacketization level $K\binom{K-(r-1)(t+L)-1}{t+L-1}$. This completes the proof of Theorem~\ref{thm_general}.
\end{IEEEproof}

\begin{rem}
	When $L=1$, the scheme in Theorem~\ref{thm_general} reduces to a single-antenna cyclic wrap-around MACC scheme having $T_n = \frac{K-rt}{t+1}$ and a subpacketization level $K\binom{K-(r-1)(t+1)-1}{t}$. The above subpacketization level requirement is lower than the scheme in \cite{CWLZC} having identical $T_n$.
\end{rem}

We now present an example to describe the scheme in Theorem~\ref{thm_general}.

\begin{figure*}
	\scriptsize{	\begin{equation}
		\mathbf{C} = 
		\begin{blockarray}{cccccccccc}
		& 1 & 2 & 3 & 4 & 5 & 6 & 7 & 8 & 9 \\
		\begin{block}{c[ccccccccc]}
		(1357,13) & \star &  & \star &  &  &  &  &  &  \\
		(1357,35) & &  & \star &  & \star &  &  &  &  \\
		(1357,57) &	&  &  &  & \star &  & \star &  &  \\
		(1357,71) & \star &  &  &  &  &  & \star  &  & \\
		(1358,13) & \star &  & \star &  &  &  &  &  &  \\
		(1358,35) & &  & \star &  & \star & &  &  &  \\
		(1358,58) & &  &  &  & \star &  &  & \star &  \\
		(1358,81) & \star &  &  &  &  &  &  & \star &  \\
		(1368,13) &	\star &  & \star & &  &  &  &  &  \\
		(1368,36) &	&  & \star &  &  & \star &  &  &  \\
		(1368,68) &	&  &  &  &  & \star &  & \star & \\
		(1368,81) &	\star &  &  &  &  &  &  & \star &   \\
		(1468,14) & \star &  &  & \star &  &  &  &  &  \\
		(1468,46) & &  &  & \star &  & \star & &  &  \\
		(1468,68) & &  &  &  &  & \star &  & \star &  \\
		(1468,81) & \star &  &  &  &  &  &  & \star &  \\
		(2468,24) & & \star &  & \star &  &  &  &  &  \\
		(2468,46) & &  &  & \star &  & \star &  &  &  \\
		(2468,68) & &  &  &  &  & \star &  & \star & \\
		(2468,82) & & \star &  &  &  &  &  & \star & \\
		(2469,24) & & \star &  & \star &  &  &  &  &  \\ 
		(2469,46) & &  &  & \star &  & \star &  &  &  \\
		(2469,69) & &  &  &  &  & \star &  & & \star  \\
		(2469,92) & & \star &  &  &  &  &  &  & \star  \\
		(2479,24) & & \star &  & \star &  &  &  &  &  \\
		(2479,47) & &  &  & \star &  &  & \star & &  \\
		(2479,79) & &  &  &  &  &  & \star &  & \star \\
		(2479,92) & & \star &  &  &  &  &  &  & \star \\
		(2579,25) &	& \star &  &  & \star &  &  &  &  \\
		(2579,57) & &  &  &  & \star &  & \star &  & \\
		(2579,79) & &  &  &  &  &  & \star &  & \star \\
		(2579,92) &	& \star &  &  &  &  &  &  & \star \\
		(3579,35) & &  & \star &  & \star &  &  &  & \\
		(3579,57) & &  &  &  & \star &  & \star &  &  \\
		(3579,79) & &  &  &  &  &  & \star &  & \star \\
		(3579,93) & &  & \star  &  &  &  &  &  & \star\\
		\end{block} \end{blockarray} \qquad
		\mathbf{D} = \left[\begin{array}{ccccccccc}
		\star & \star & \star & 4 & 2 & 3 & 1 & 10 & \star \\
		1 & \star & \star & \star & \star & 4 & 2 & 25 & 3 \\
		2 & 3 & 1 & \star & \star & \star & \star & 20 & 4 \\
		\star & 4 & 2 & 3 & 1 & \star & \star  & 15 & \star \\
		\star & \star & \star & 9 & 7 & 30 & 8 & 6 & \star \\
		6 & \star & \star & \star & \star & 25 & 9 & 7 & 8 \\
		7 & 8 & 6 & \star & \star & 20 & \star & \star & 9 \\
		\star & 9 & 7 & 8 & 6 & 15 & \star & \star & \star \\
		\star & \star & \star & 30 & 14 & 12 & 13 & 11 & \star \\
		11 & \star & \star & 25 & \star & \star & 14 & 12 & 13 \\
		12 & 13 & 11 & 20 & \star & \star & \star & \star & 14 \\
		\star & 14 & 12 & 35 & 13 & 11 & \star & \star & \star  \\
		\star & 30 & \star & \star & 19 & 17 & 18 & 16 & \star \\
		16 & 25 & \star & \star & \star & \star & 19 & 17 & 18 \\
		17 & 40 & 18 & 16 & \star & \star & \star & \star & 19 \\
		\star & 35 & 19 & 17 & 18 & 16 & \star & \star & \star \\
		\star & \star & \star & \star & 24 & 22 & 23 & 21 & 30 \\
		23 & 21 & \star & \star & \star & \star & 24 & 22 & 45 \\
		24 & 22 & 23 & 21 & \star & \star & \star  & \star & 40 \\
		\star & \star &  24 & 22 & 23 & 21 & \star & \star & 35 \\
		\star & \star & \star & \star & 29 & 27 & 5 & 28 & 26 \\ 
		28 & 26 & \star & \star & \star & \star & 45 & 29 & 27 \\
		29 & 27 & 28 & 26 & \star & \star & 40 & \star & \star \\
		\star & \star &  29 & 27 & 28 & 26 & 35 & \star & \star \\
		\star & \star & \star & \star & 5 & 34 & 32 & 33 & 31 \\
		33 & 31 & \star & \star & 45 & \star & \star & 34 & 32 \\
		34 & 32 & 33 & 31 & 40 & \star & \star & \star & \star \\
		\star & \star & 34 & 32 & 10 & 33 & 31 & \star & \star \\
		\star & \star & 5 & \star & \star & 39 & 37 & 38 & 36 \\
		38 & 36 & 45 & \star & \star & \star & \star & 39 & 37 \\
		39 & 37 & 15 & 38 & 36 & \star & \star & \star & \star \\
		\star & \star & 10 & 39 & 37 & 38 & 36 & \star & \star \\
		5 & \star & \star & \star & \star & 44 & 42 & 43 & 41 \\
		20 & 43 & 41 & \star & \star & \star & \star & 44 & 42 \\
		15 & 44 & 42 & 43 & 41 & \star & \star & \star & \star \\
		10 & \star & \star  & 44 & 42 & 43 & 41 & \star & \star
		\end{array}\right]
		\label{eq:ex_matrix}
		\end{equation}}
	\hrule
\end{figure*}
\begin{exmp}
 Consider a cyclic wrap-around MACC network with a server having $N=9$ unit-sized files $W^{[9]}=\{W^1,W^2,\ldots,W^9\}$ and $L=2$ transmit antennas. There are $K=9$ users and caches, and each user $k \in [9]$ accesses $r=2$ consecutive caches $C_k$ and $C_{\langle k+1 \rangle_9}$. Each cache is of size $M=2$ units, thus obtaining $t=2$. The condition $K \geq r(t+L)$ holds in this case.
\end{exmp}

We first construct a $(9,36,8,2)$ caching array $\mathbf{C}$ and a $(\mathbf{C},45,2)$ delivery array $\mathbf{D}$ according to Algorithm~\ref{alg_const}. The columns of $\mathbf{C}$ and $\mathbf{D}$ are indexed using the set $[9]$, and the rows of $\mathbf{C}$ and $\mathbf{D}$ are indexed using the elements in $\mathcal{I}$, which is given below.
\begin{equation*}
\begin{aligned}
\small
& \mathcal{I}=\{(1357,13),(1357,35),(1357,57),(1357,71),\\
&  (1358,13), (1358,35), (1358,58), (1358,81), (1368,13),\\
& (1368,36),(1368,68),(1368,81), (1468,14),(1468,46),\\
& (1468,68), (1468,81),(2468,24),(2468,46),(2468,68),\\
&(2468,82),(2469,24),(2469,46),(2469,69),(2469,92) \\
& (2479,24),(2479,47),(2479,79),(2479,92),(2579,25),\\
& (2579,57),(2579,79),(2579,92),(3579,35), (3579,57),\\
& (3579,79),(3579,93)\}.
\end{aligned}
\end{equation*}
The $(9,36,8,2)$ caching array $\mathbf{C}$ and the $(\mathbf{C},45,2)$ delivery array $\mathbf{D}$ are given in \eqref{eq:ex_matrix}. Each file $W^n, n \in [9]$, is divided into $36$ non-overlapping equal-sized subfiles, where each subfile is indexed using a two tuple $(\mathcal{J},w_{\mathcal{J}}) \in \mathcal{I}$. The $i^{th}$ column of $\mathbf{C}$ represents cache $C_i$. The content placement in each cache is determined by the $\star$'s in the corresponding column of $\mathbf{C}$, and the contents stored in each cache are as follows:
\begin{align*}
 \mathcal{Z}_{C_1} & = \{W^n_{(1357,13)}, W^n_{(1357,71)}, W^n_{(1358,13)}, W^n_{(1358,81)}, \\
 & W^n_{(1368,13)},  W^n_{(1368,81)}, W^n_{(1468,14)}, W^n_{(1468,81)}, \forall n \in [9]\},\\
 \mathcal{Z}_{C_2} & =\{W^n_{(2468,24)},W^n_{(2468,82)},W^n_{(2469,24)},W^n_{(2469,92)},\\
 & W^n_{(2479,24)},W^n_{(2479,92)},W^n_{(2579,25)},W^n_{(2579,92)},\forall n \in [9]\}, \\
 \mathcal{Z}_{C_3} & = \{W^n_{(1357,13)}, W^n_{(1357,35)}, W^n_{(1358,13)}, W^n_{(1358,35)}, \\
 & W^n_{(1368,13)},  W^n_{(1368,36)}, W^n_{(3579,35)}, W^n_{(3579,93)}, \forall n \in [9]\},\\
  \mathcal{Z}_{C_4} & =\{W^n_{(1468,14)}, W^n_{(1468,46)},W^n_{(2468,24)},W^n_{(2468,46)},\\ & W^n_{(2469,24)},W^n_{(2469,46)}, W^n_{(2479,24)},W^n_{(2479,47)},\forall n \in [9]\}, \\
  \mathcal{Z}_{C_5} & = \{W^n_{(1357,35)}, W^n_{(1357,57)}, W^n_{(1358,35)}, W^n_{(1358,58)}, \\
  &  W^n_{(2579,25)},W^n_{(2579,57)},W^n_{(3579,35)}, W^n_{(3579,57)}, \forall n \in [9]\},\\
   \mathcal{Z}_{C_6} & =\{W^n_{(1368,36)},  W^n_{(1368,68)}, W^n_{(1468,46)}, W^n_{(1468,68)},\\
  & W^n_{(2468,46)},W^n_{(2468,68)},W^n_{(2469,46)},W^n_{(2469,69)}, \forall n \in [9]\}, \\
   \mathcal{Z}_{C_7} & = \{W^n_{(1357,57)}, W^n_{(1357,71)}, W^n_{(2479,47)},W^n_{(2479,79)}, \\
  &  W^n_{(2579,25)},W^n_{(2579,57)},W^n_{(3579,35)}, W^n_{(3579,57)}, \forall n \in [9]\},\\
  \mathcal{Z}_{C_8} & =\{ W^n_{(1358,13)}, W^n_{(1358,81)},W^n_{(1368,68)},  W^n_{(1368,81)}, \\
  & W^n_{(1468,68)}, W^n_{(1468,81)}, W^n_{(2468,68)},W^n_{(2468,82)}, \forall n \in [9]\}, \\
  \mathcal{Z}_{C_9} & = \{W^n_{(2469,69)},W^n_{(2469,92)},W^n_{(2479,79)},W^n_{(2479,92)}, \\
  &  W^n_{(2579,79)},W^n_{(2579,92)},W^n_{(3579,79)}, W^n_{(3579,93)}, \forall n \in [9]\}.
 \end{align*}
Each user $k \in [9]$ has access to the caches $C_k$ and $C_{\langle k+1 \rangle_9}$. Hence, each user gets $rZ=16$ subfiles of every file $W^n, n \in [9]$. The subfiles known to each user are represented by $\star$'s in $\mathbf{D}$. Assume that the demand vector is $\mathbf{d}=(1,2,3,4,5,6,7,8,9)$. There are transmissions corresponding to every distinct integer in $\mathbf{D}$, and each transmission benefits $t+L=4$ users. In the interest of space, we list only three transmissions. The transmissions corresponding to $s=1$, $s=4$, and $s=5$ are given below.
\begin{align*}
   \mathbf{x}^{(1)} &= \mathbf{V}^{(1)}[W^1_{(1357,35)},W^3_{(1357,57)},W^5_{(1357,71)]}, W^7_{(1357,13)}]^{\mathrm{T}} \\
   \mathbf{x}^{(4)} & = \mathbf{V}^{(4)}[W^2_{(1357,71)},W^4_{(1357,13)},W^6_{(1357,35)},W^9_{(1357,57)}]^{\mathrm{T}}\\
   \mathbf{x}^{(5)} & = \mathbf{V}^{(5)}[W^1_{(3579,35)},W^3_{(2579,25)},W^5_{(2479,24))},W^7_{(2469,24)}]^{\mathrm{T}}
\end{align*}
{Note that the transmissions $\mathbf{x}^{(1)}$ and $\mathbf{x}^{(5)}$ benefit users $1$, $3$, $5$, and $7$, and each of them gets one subfile of its desired file from each of the two transmissions.} Similarly,  $\mathbf{x}^{(4)}$ serves users $2$, $4$, $6$, and $7$. To explain the decoding, consider user 1 and transmission $\mathbf{x}^{(1)}$. Let ${y}^{(1)}_1$ be the received message at user $1$ corresponding to $\mathbf{x}^{(1)}$. Then, ${y}^{(1)}_1$ can be written as follows:
\begin{subequations}
	\begin{align}
  {y}^{(1)}_1 & =\mathbf{h}_1^{\mathrm{T}}\mathbf{x}^{(1)}+n_1 \label{eq:ex_gen1}\\
  & = \mathbf{h}_1^{\mathrm{T}}[\mathbf{v}_1^{(1)} \textrm{\hspace{0.1cm}}\mathbf{v}_3^{(1)}\textrm{\hspace{0.1cm}}\mathbf{v}_5^{(1)}\textrm{\hspace{0.1cm}}\mathbf{v}_7^{(1)}]\begin{bmatrix}
  W^1_{(1357,35)}\\
  W^3_{(1357,57)} \\
  W^5_{(1357,71)}\\
  W^7_{(1357,13)}
  \end{bmatrix} \label{eq:ex_gen2}\\
 & = \mathbf{h}_1^{\mathrm{T}}\mathbf{v}_1^{(1)} W^1_{(1357,35)} + \mathbf{h}_1^{\mathrm{T}}\mathbf{v}_5^{(1)} W^5_{(1357,71)}+\notag\\
 & \textrm{\hspace{4.5cm}}\mathbf{h}_1^{\mathrm{T}}\mathbf{v}_7^{(1)}W^7_{(1357,13)}.
 \label{eq:ex_gen3}
  \end{align}
\end{subequations}

The additive white gaussian noise $n_1$ is neglected in \eqref{eq:ex_gen2} due to high SNR assumption. The design of $\mathbf{v}_3^{(1)} \in \mathbb{C}^{2 \times 1}$ ensures that $\mathbf{h}_1^{\mathrm{T}}\mathbf{v}_3^{(1)}=0$. In \eqref{eq:ex_gen3}, the terms $\mathbf{h}_1^{\mathrm{T}}\mathbf{v}_5^{(1)} W^5_{(1357,71)}$ and $\mathbf{h}_1^{\mathrm{T}}\mathbf{v}_7^{(1)}W^7_{(1357,13)}$ can be eliminated as the channel coefficients and the subfiles $W^5_{(1357,71)}$ and $W^7_{(1357,13)}$ are known to user $1$. Thus, user $1$ gets the subfile $W^1_{(1357,35)}$. In a similar manner, user $1$ decodes all the required subfiles from the transmissions. The decoding procedure remains the same for all the users. Thus, the normalized delivery time is obtained as $T_n = 45/36=9/4$.

\subsubsection{An approach to reduce subpacketization level when $\gcd(K,t,L)\neq 1$}
\label{subsubsec:subredn}
The multi-antenna MACC scheme obtained from Construction I requires a subpacketization level $F = K\binom{K-(r-1(t+L)-1)}{t+L-1}$. This $F$ can be further reduced if $\gcd(K,t,L) \neq 1$. Let $\gamma = \gcd(K,t,L)$ and consider a multi-antenna MACC network with parameters $K,r,t,$ and $L$ such that $\gamma \neq 1$. Then, let $K^{\prime} = \frac{K}{\gamma}$, $t^{\prime}=\frac{t}{\gamma}$, and $L^{\prime}=\frac{L}{\gamma}$. Using $K^{\prime},r,t^{\prime}$, and $L^{\prime}$ as the new parameters, we construct a $(K^{\prime},K^{\prime}\binom{K^{\prime}-(r-1)(t^{\prime}+L^{\prime})-1}{t^{\prime}+L^{\prime}-1}$, $t^{\prime}\binom{K^{\prime}-(r-1)(t^{\prime}+L^{\prime})-1}{t^{\prime}+L^{\prime}-1},r)$ caching array $\mathbf{C}^{\prime}$ as in Algorithm~\ref{alg_const}. Then, $\mathbf{C}^{\prime}$ is concatenated horizontally $\gamma-$times to obtain an array $\mathbf{C}$ of size $K^{\prime}\binom{K^{\prime}-(r-1)(t^{\prime}+L^{\prime})-1}{t^{\prime}+L^{\prime}-1} \times K$. It is straightforward that $\mathbf{C}$ satisfies the condition $B1$. The array $\mathbf{C}$ also satisfies condition $B2$ by virtue of the $\gamma-$times horizontal concatenation of the caching array $\mathbf{C}^{\prime}$. In $\mathbf{C}$, the set of rows containing $\star$'s in each of the following columns $k,k+K^{\prime},\ldots,k + (\gamma-1)K^{\prime}$ is the same, where $k \in [K^{\prime}]$ . Thus, the horizontal concatenation of $\mathbf{C}^{\prime}$ results in a $(K,K^{\prime}\binom{K^{\prime}-(r-1)(t^{\prime}+L^{\prime})-1}{t^{\prime}+L^{\prime}-1},t^{\prime}\binom{K^{\prime}-(r-1)(t^{\prime}+L^{\prime})-1}{t^{\prime}+L^{\prime}-1},r)$ caching array $\mathbf{C}$. The cache placement is done using the $(K,K^{\prime}\binom{K^{\prime}-(r-1)(t^{\prime}+L^{\prime})-1}{t^{\prime}+L^{\prime}-1},t^{\prime}\binom{K^{\prime}-(r-1)(t^{\prime}+L^{\prime})-1}{t^{\prime}+L^{\prime}-1},r)$ caching array $\mathbf{C}$, where $\mathbf{C}$ meets the condition $\frac{Z}{F}=\frac{t^{\prime}}{K^{\prime}}=\frac{M}{N}$. Notice that for any $k \in [K]$, the set of caches $\{\langle k+(i-1)K^{\prime}\rangle_K, \forall i \in [\gamma]\}$ store exactly the same contents.
 
 To construct a $(\mathbf{C},r,S,L)$ delivery array $\mathbf{D}$, we first construct a $(\mathbf{C}^{\prime},r,\frac{K^{\prime}(K^{\prime}-rt^{\prime})}{t^{\prime}+L^{\prime}}\binom{K^{\prime}-(r-1)(t^{\prime}+L^{\prime})-1}{t^{\prime}+L^{\prime}-1},L^{\prime})$ delivery array $\mathbf{D}^{\prime}$ as in Algorithm~\ref{alg_const}. Then, $\mathbf{D}^{\prime}$ is concatenated $\gamma-$times horizontally to get another array $\mathbf{D}$ of size $K^{\prime}\binom{K^{\prime}-(r-1)(t^{\prime}+L^{\prime})-1}{t^{\prime}+L^{\prime}-1} \times K$. The number of integers in $\mathbf{D}$ and $\mathbf{D}^{\prime}$ is the same, but the occurrence of each integer in $\mathbf{D}$ is $\gamma-$times more than that of $\mathbf{D}^{\prime}$. The array $\mathbf{D}$ is, in fact, a $(\mathbf{C}, r, S,L)$ delivery array. It is easy to see that the position of $\star$'s in $\mathbf{D}$ is according to the caching array $\mathbf{C}$, thus condition $D1$ is satisfied. Condition $D2$ directly follows from $\mathbf{D}^{\prime}$ as it is a $(\mathbf{C}^{\prime},\frac{K^{\prime}(K^{\prime}-rt^{\prime})}{t^{\prime}+L^{\prime}}\binom{K^{\prime}-(r-1)(t^{\prime}+L^{\prime})-1}{t^{\prime}+L^{\prime}-1},L^{\prime})$ delivery array. Now, it remains to verify condition $D3$. Let $S=\frac{K^{\prime}(K^{\prime}-rt^{\prime})}{t^{\prime}+L^{\prime}}\binom{K^{\prime}-(r-1)(t^{\prime}+L^{\prime})-1}{t^{\prime}+L^{\prime}-1}$ and consider any integer $s \in [S]$ in $\mathbf{D}$. Then, any row in $\mathbf{D}^{(s)}$ contains only $L$ integers at the most because $\mathbf{D}^{(s)}=\underbrace{[\mathbf{D}^{\prime^{(s)}},\ldots,\mathbf{D}^{\prime{(s)}}]}_\text{$\gamma-$times}$ and each row of $\mathbf{D}^{\prime^{(s)}}$ has only at most $L^{\prime}$ integers. Thus, array $\mathbf{D}$ satisfies $D1, D2,$ and $D3$, and hence is a $(\mathbf{C},\frac{K^{\prime}(K^{\prime}-rt^{\prime})}{t^{\prime}+L^{\prime}}\binom{K^{\prime}-(r-1)(t^{\prime}+L^{\prime})-1}{t^{\prime}+L^{\prime}-1},L)$ delivery array. Corresponding to a demand vector $\mathbf{d}$, there are $S=\frac{K^{\prime}(K^{\prime}-rt^{\prime})}{t^{\prime}+L^{\prime}}\binom{K^{\prime}-(r-1)(t^{\prime}+L^{\prime})-1}{t^{\prime}+L^{\prime}-1}$ transmissions, each transmission takes a normalized delivery time of ${1}/{K^{\prime}\binom{K^{\prime}-(r-1)(t^{\prime}+L^{\prime})-1}{t^{\prime}+L^{\prime}-1}}$. Thus, the normalized delivery is obtained as $T_n = \frac{K^{\prime}-rt^{\prime}}{t^{\prime}+L^{\prime}}=\frac{K-rt}{t+L}$. In a nutshell, we show that the subpacketization level of the scheme in Theorem~\ref{thm_general} can be reduced from $F = K\binom{K-(r-1(t+L)-1)}{t+L-1}$ to $F^{\prime}={K^{\prime}\binom{K^{\prime}-(r-1)(t^{\prime}+L^{\prime})-1}{t^{\prime}+L^{\prime}-1}}$ without affecting the normalized delivery time when $\gcd(K,t,L) \neq 1$.
 
 To see the amount of reduction achieved in the subpacketization level by the above approach, consider an example with the following parameters: $N=20$, $K=20$, $r=3$, $t=4$, and $L=2$. We just focus on the subpacketization level required in the two approaches and do not delve into the detailed description of the schemes. If we directly follow  the scheme in Theorem~\ref{thm_general}, the subpacketization level obtained is $F = K \binom{K-(r-1)(t+L)-1}{t+L-1}= 480$. Since $\gcd(K,t,L)=2$, we get $F^{\prime}={K^{\prime}\binom{K^{\prime}-(r-1)(t^{\prime}+L^{\prime})-1}{t^{\prime}+L^{\prime}-1}}=30$ which is far less than $F$.
\subsection{Constructions II and III} 
\label{subsec:const2_3}
Next, we present two constructions of caching and delivery arrays that result in multi-antenna MACC schemes achieving optimal performance, under uncoded placement and one-shot delivery, in some instances. To achieve this optimal performance, each integer should appear $rt+L$ times  in the delivery array.

 \begin{thm}
	For a cyclic wrap-around MACC network, the normalized delivery time 
	\begin{equation}
	T_n=\frac{K-rt}{rt+L} 
	\label{eq:rate} 
	\end{equation}
	is achievable in the following cases: \\
	\textit{(a)} $K=rt+L$ and $\gcd(K,t)=1$, \\
	\textit{(b)} $K=mrt+(m-1)L$, $m \in \mathbb{Z}$, $m \geq 2 $, $L \geq rt$, and $\gcd(K,t)=1$.\\
	Furthermore, the normalized delivery time in \eqref{eq:rate} is optimal under uncoded placement and one-shot delivery.
	\label{thm1}
\end{thm}

 \begin{IEEEproof} 
	\textit{Construction II.}  Consider case (a) $K=rt+L$ and $\gcd(K,t)=1$. We need to construct a $(K,K,t,r)$ caching array $\mathbf{C}$ and a $(\mathbf{C},L,L)$ delivery array $\mathbf{D}$ for this case.
	 
 	The caching array $\mathbf{C}$ is of size $K \times K$, and there are $t$ $\star$'s present in each column of $\mathbf{C}$.
 	The position of $\star$'s in the $k^{th}$ column of $\mathbf{C}$ is given by:
 	\begin{equation}
 	   c_{j,k} = \star, \forall j \in [\langle (k-1)t+1 \rangle_K, \langle kt \rangle_K].
 	   \label{eq:star_plac}
 	\end{equation}
 \noindent The  placement in \eqref{eq:star_plac} satisfies the condition $B2$ as $rt < K$. Next, we construct a $(\mathbf{C},L,L)$ delivery array $\mathbf{D}$ of size $K \times K$. The position of $\star$'s in the $k^{th}$ column of $\mathbf{D}$ is given by:
 	\begin{equation}
 	   d_{j,k}=\star, \forall j \in \cup_{i \in [k,\langle k+r-1 \rangle_K]} [\langle(i-1)t+1\rangle_K,\langle it \rangle_K].
 	   \label{eq:user_access}
 	\end{equation}
 	Thus, each column of $\mathbf{D}$ has $rt$ number of $\star$'s by \eqref{eq:star_plac} and \eqref{eq:user_access}. {Notice that in each column, the $rt$ $\star$'s appear consecutively. According to \eqref{eq:user_access}, the $\star$'s in the $k^{th}$ column of $\mathbf{D}$ start appearing consecutively from the $\langle (k-1)t+1 \rangle_K^{th}$ row in a cyclic wrap-around manner. Therefore, the  set $\{1, t+1,\langle 2t+1 \rangle_K,\ldots, \langle (K-1)t+1 \rangle_K \}$ represents the smallest row index from which $\star$'s start appearing consecutively in the columns $\{1,2,\ldots,K\}$ of $\mathbf{D}$, respectively. Next, we perform a column permutation on $\mathbf{D}$ to obtain $\mathbf{D}^{\prime}=[\mathbf{d}_1^{\prime}, \mathbf{d}_2^{\prime}, \ldots, \mathbf{d}_K^{\prime}]$, where the position of $\star$'s in any column $k \in [K]$ of $\mathbf{D}^{\prime}$ is given by:  $d^{\prime}_{j,k}=\star$, $\forall j \in [k, \langle k+rt-1 \rangle_K ]$. i.e., in column $\mathbf{d}^{\prime}_k$, the $\star$'s appear consecutively from the $k^{th}$ row. Since $\gcd(K,t)=1$, there exists a permutation on the columns of $\mathbf{D}$ that leads to $\mathbf{D}^{\prime}$. Let $\pi$ be the function that performs the above permutation on the columns of $\mathbf{D}$ and results in $\mathbf{D}^{\prime}$. Therefore in $\mathbf{D}^{\prime}$, the column $\mathbf{d}_1^{\prime}$ has $\star$'s present in the first $rt$ rows, the column $\mathbf{d}_2^{\prime}$ has $\star$'s from second row to $rt+1$, and the position of $\star$'s in any $k^{th}$ column is obtained by a unit cyclic shift of the position of $\star$'s in the $\langle k-1\rangle_K^{th}$ column downwards.
    If $t=1$, $\pi$ is an identity function which implies $\mathbf{D}^{\prime}=\mathbf{D}$.} The column permutation is done to simplify the integer assignment procedure.
    
    The next step is to fill integers in $K(K-rt)=KL$ vacant cells of $\mathbf{D}^{\prime}$ by satisfying conditions \textit{D2} and \textit{D3}. By condition \textit{D2}, at least $L$ distinct integers are required to fill $\mathbf{D}^{\prime}$, and we show that it is possible to do with exactly $L$ integers. Consider an integer $s \in [L]$. Then, we have
    \begin{equation}
      d_{j,k}^{\prime}=s, \forall k \in [K] \textrm{ and } j = \langle rt+s+k-1 \rangle_K.
      \label{eq:int_pla}
    \end{equation}
 	From \eqref{eq:int_pla}, it is evident that each integer appears $K$ times in $\mathbf{D}^{\prime}$, i.e, once in every column, thus, guaranteeing condition \textit{D2}. Now, it remains to verify condition \textit{D3}. Since each integer appears once in every column and row, $\mathbf{D}^{\prime (s)}=\mathbf{D}^{\prime}$. Note that, by construction, each row of $\mathbf{D}^{\prime}$ also has $rt$ number of $\star$'s, which implies there are only $K-rt=L$ integers present in any row of $\mathbf{D}^{\prime}$. Hence, condition \textit{D3} is satisfied. Now, we apply the inverse function $\pi^{-1}$ on the columns of $\mathbf{D}^{\prime}$, and obtain $\mathbf{D}$ which is a $(\mathbf{C},L,L)$ delivery array. The conditions \textit{D2} and \textit{D3} are unaffected by the column permutations. In fact, we do not need to perform the column permutations in this case. The $L$ integers can be placed in any random order in a column. The condition \textit{D3} is still valid as the number of integers present in a row is always $L$.

	Now, using Lemma~\ref{ref:lem}, the $(K,K,t,r)$ caching array $\mathbf{C}$ and the $(\mathbf{C},L,L)$ delivery array $\mathbf{D}$ together result in a multi-antenna MACC scheme with a normalized delivery time
	\begin{equation}
	  T_n= \frac{L}{K}= \frac{K-rt}{rt+L}
	  \label{eq:rate_th}
	\end{equation}
 	and a subpacketization level $K$.  From \eqref{eq:bound}, we get $T_n = T^{*}_{n,u}=\frac{K-rt}{rt+L}$. Thus, $T_n$ in \eqref{eq:rate_th} is optimal.
 	
 	\textit{Construction III. } Next, consider case (b) $K=mrt+(m-1)L$, where $L \geq rt$, $\gcd(K,t)=1$, and $m$ is an integer such that $m \geq 2$. In this case, we need to construct a $(K,K,t,r)$ caching array $\mathbf{C}$ and a $(\mathbf{C},(m-1)K,L)$ delivery array $\mathbf{D}$. The construction of $\mathbf{C}$ is same as in case (a) and follows \eqref{eq:star_plac}. Therefore, the position of $\star$'s in $\mathbf{D}$ is also according to \eqref{eq:user_access}. Thus, each column of $\mathbf{D}$ contains $rt$ number of $\star$'s, and we permute the columns of $\mathbf{D}$ using $\pi$ as done in case $(a)$ to obtain $\mathbf{D}^{\prime}=[\mathbf{d}^{\prime}_1,\mathbf{d}^{\prime}_2,\ldots,\mathbf{d}^{\prime}_K]$. In $\mathbf{D}^{\prime}$, the $\star$'s in the column $\mathbf{d}^{\prime}_k$, $k \in [K]$, appears from $k^{th}$ row to $\langle k+rt-1 \rangle_K^{th}$ row consecutively. This arrangement of $\star$'s is required to follow a specific integer assignment. There are $K(K-rt)$ vacant cells in $\mathbf{D}^{\prime}$, and they are filled using $(m-1)K$ integers satisfying conditions \textit{D2} and \textit{D3}. Each integer $s \in [(m-1)K]$ occurs $rt+L$ times in $\mathbf{D}^{\prime}$. Consider any integer $s \in [(m-1)K]$, where $s=pK+q$, $p \in [0,m-2]$, $q \in [K]$. If $p$ is even, the integer $s$ occurs $L$ times in the $q^{th}$ row and $rt$ times in the $\langle \frac{p}{2}(rt+L)+rt+q \rangle_K^{th}$ row of $\mathbf{D}^{\prime}$ as given below:
 	\begin{subequations}
 	\begin{align}
 	  d^{\prime}_{q,\langle \frac{p}{2}(rt+L)+q+i\rangle_K}&=s \textrm{\hspace{0.5cm}}\forall i \in [L], \label{eq:s_evena}\\
 	  d^{\prime}_{\langle \frac{p}{2}(rt+L)+rt+q \rangle_K,\langle q-rt+i \rangle_K} &=s \textrm{\hspace{0.5cm}}\forall i \in [rt].
 	  \label{eq:s_evenb}
 	  \end{align}
 	\end{subequations}
 	If $p$ is odd, the integer $s$ occurs $rt$ times in $q^{th}$ row, and $L$ times in the $\langle (\frac{p+1}{2})(rt+L)+rt+q\rangle_K^{th}$ row as follows:
 	\begin{subequations}
 	\begin{align}
 	  d^{\prime}_{q,\langle(\frac{p-1}{2})(rt+L)+L+q+i\rangle_K}&=s \textrm{\hspace{0.5cm}}\forall i \in [rt], \label{eq:s_odda}\\
 	  d^{\prime}_{\langle(\frac{p+1}{2})(rt+L)+q\rangle_K,\langle q-rt+i\rangle_K}&=s \textrm{\hspace{0.5cm}}\forall i \in [L].
 	  \label{eq:s_oddb}
 	\end{align}
	\end{subequations}
 	Note that the above assignment of integers in $\mathbf{D}^{\prime}$ is according to an EPDA construction (Construction II) in \cite{NPR}. Therefore, conditions \textit{D2} and \textit{D3} follow directly from the EPDA conditions.
 	
 	 Now, undo the permutation on $\mathbf{D}^{\prime}$ using $\pi^{-1}$ to obtain $\mathbf{D}$. The conditions \textit{D2} and \textit{D3} hold even under column rearrangement. Thus, we obtained a $(\mathbf{C},(m-1)K,L)$ delivery array $\mathbf{D}$. Then, using $\mathbf{C}$ and $\mathbf{D}$, a multi-antenna MACC scheme can be obtained with a normalized delivery time $T_n=(m-1)K/K=(m-1)$ and a subpacketization level $K$. From the condition $K=mrt+(m-1)L$, we get $(m-1)=\frac{K-rt}{rt+L}$. Thus, $T_n=\frac{K-rt}{rt+L}$ and from \eqref{eq:bound}, we know that it is optimal under uncoded placement and one-shot delivery. This completes the proof of Theorem~\ref{thm1}.
 	\end{IEEEproof}
 	Constructions II and III are based on the two EPDA constructions presented in \cite{NPR}.  The advantage of these two constructions is that the resulting schemes achieve optimal performance with a subpacketization level that is linear with respect to the number of  users. The applicability of the schemes in Theorem \ref{thm1} is limited but  there exist scenarios where the network parameters satisfy any of the conditions in case (a) or case (b). For instance, consider a network with a large number of users, moderate cache size, and a smaller value for $r$. In such a scenario, if the number of transmit antennas at the server can be made large, it is possible to satisfy the condition given in case (b) in Theorem~\ref{thm1}. Similarly, for a given $K$, $r$, and $t$, the condition
 	$K = rt + L$ in case (a) can be satisfied by selecting an appropriate $L$, the number of transmit antennas with the server.
 	
 \begin{rem}
 When $L=1$ in case (a) $K=rt+L$ with $\gcd(K,t)=1$, the proposed scheme recovers the single-antenna MACC scheme  in \cite{SPE} which is optimal under uncoded placement. In \cite{SPE}, the authors  considered a case where $r=\frac{K-1}{t}$.
 \label{remark1}
\end{rem}

 
 \begin{figure*}[t!]
 	\centering
 	\includegraphics[width=0.9\textwidth]{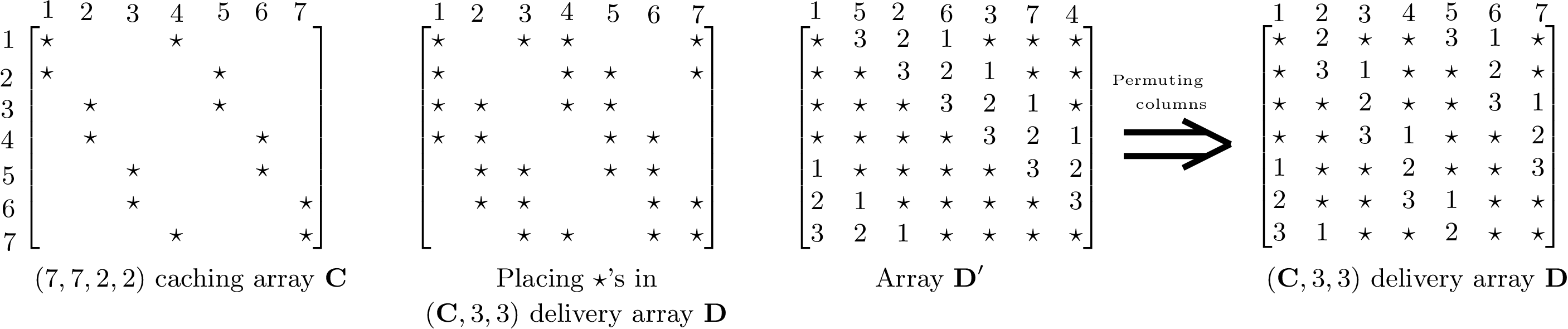}
 	\caption{Caching and delivery arrays for $K=7$, $N=7$, $r=2$, $L=3$, $t=2$.}
 	\label{fig:example_t2}
 \end{figure*}

\begin{figure*}[t!]
	\centering
	\includegraphics[width=0.7\textwidth]{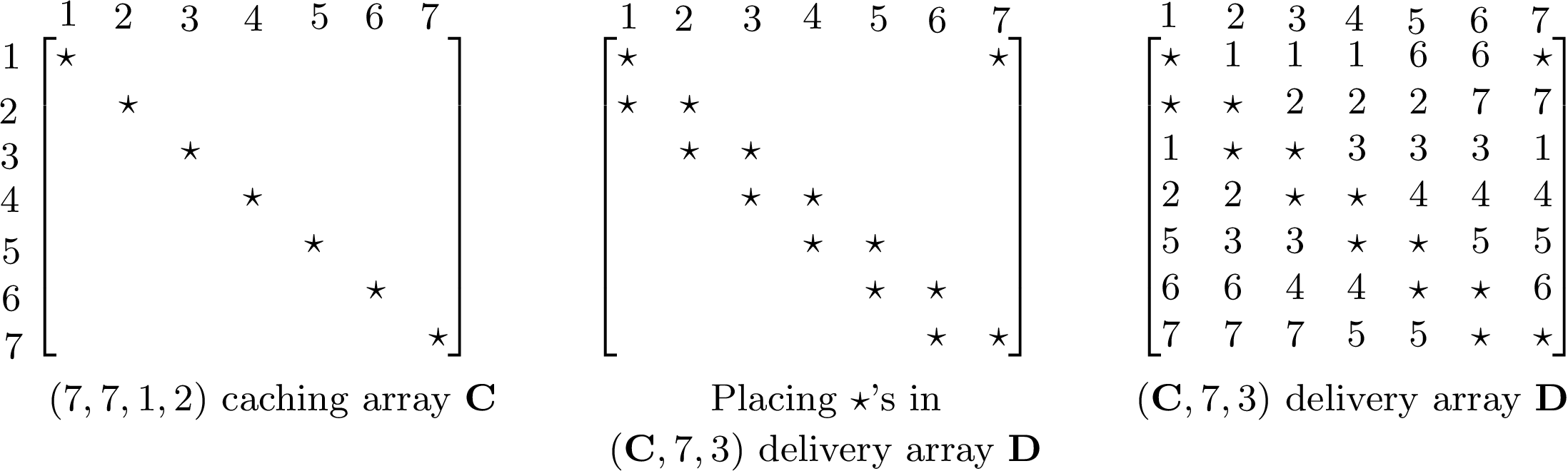}
	\caption{Caching and delivery arrays for $K=7$, $N=7$, $r=2$, $L=3$, $t=1$.}
	\label{fig:example_t1}
\end{figure*}
 
  \begin{exmp}
 	Consider a MACC network with a server having $L=3$ transmit antennas and $N=7$ unit-sized files $W^{[7]}=\{W^1,W^2,\ldots,W^7\}$. There are $K=7$ users and as many number of caches, and each cache is of size $M=2$ files. Each user accesses $r=2$ neighboring caches, i.e., user $k$ accesses caches $C_k$ and $C_{\langle k+1 \rangle_7}$.
 	\label{ex:exmp2}
  \end{exmp}
 
 In this example, $t=2$ and the condition $K=rt+L$ is satisfied. Also, $\gcd(K,t)=1$ in this case. Therefore, we construct a $(7,7,2,2)$ caching array $\mathbf{C}$ and a $(\mathbf{C},3,3)$ delivery array $\mathbf{D}$ as shown in Fig.~\ref{fig:example_t2}. The array $\mathbf{C}$ determines the cache placement. Each file $W^n$, $n \in [7]$, is divided into $7$ subfiles: $W^n=\{W^n_1,W^n_2,\ldots,W^n_7\}$. The contents stored in each cache are as follows:
 \begin{equation*}
\begin{aligned}
  \mathcal{Z}_{C_1}&=\{W^n_1,W^n_2, \forall n \in [7]\},  \mathcal{Z}_{C_2}=\{W^n_3,W^n_4, \forall n \in [7]\},\\
  \mathcal{Z}_{C_3}&=\{W^n_5,W^n_6, \forall n \in [7]\},  \mathcal{Z}_{C_4}=\{W^n_7,W^n_1, \forall n \in [7]\},\\
  \mathcal{Z}_{C_5}&=\{W^n_2,W^n_3, \forall n \in [7]\},  \mathcal{Z}_{C_6}=\{W^n_4,W^n_5, \forall n \in [7]\},\\
  \mathcal{Z}_{C_7}&=\{W^n_6,W^n_7, \forall n \in [7]\}.
\end{aligned}
\end{equation*}
The second array in Fig.~\ref{fig:example_t2} corresponds to the contents accessible to each user. Since user $k \in [7]$ has access to caches $C_k$ and $C_{\langle k+1 \rangle_7}$, the contents known to each user are: 
\begin{equation*}
\begin{aligned}
Z_1 & = \mathcal{Z}_{\mathcal{C}_1} \cup \mathcal{Z}_{\mathcal{C}_2},
Z_2 = \mathcal{Z}_{\mathcal{C}_2} \cup \mathcal{Z}_{\mathcal{C}_3}, 
Z_3 = \mathcal{Z}_{\mathcal{C}_3} \cup \mathcal{Z}_{\mathcal{C}_4},\\
Z_4 & = \mathcal{Z}_{\mathcal{C}_4} \cup \mathcal{Z}_{\mathcal{C}_5}, 
Z_5 = \mathcal{Z}_{\mathcal{C}_5} \cup \mathcal{Z}_{\mathcal{C}_6},
Z_6 = \mathcal{Z}_{\mathcal{C}_6} \cup \mathcal{Z}_{\mathcal{C}_7}, \\
Z_7 & = \mathcal{Z}_{\mathcal{C}_7} \cup \mathcal{Z}_{\mathcal{C}_1}.
\end{aligned}
\end{equation*}
Before filling the integers in $\mathbf{D}$, the columns are permuted according to a map $\mathbf{\pi}: \{1,2,3,4,5,6,7\} \rightarrow \{1,5,2,6,3,7,4\}$ and results in another array $\mathbf{D}^{\prime}=[\mathbf{d}_1,\mathbf{d}_5,\mathbf{d}_2,\mathbf{d}_6,\mathbf{d}_3,\mathbf{d}_7,\mathbf{d}_4]$. Once integers are filled in $\mathbf{D}^{\prime}$, the columns of $\mathbf{D}^{\prime}$ are rearranged according to the map $\pi^{-1}$. Thus, we obtained a $(\mathbf{C},3,3)$ delivery array $\mathbf{D}$. The $k^{th}$ column of $\mathbf{D}$ represents user $k$. Let the demand vector be $\mathbf{d}=(1,2,3,4,5,6,7)$. There is a transmission corresponding to each $s \in [3]$. The transmissions are:
\begin{equation*}
\begin{aligned}
  \mathbf{x}^{(1)}&=\mathbf{V^{(1)}}[W^1_5, W^2_7, W^3_2, W^4_4, W^5_6, W^6_1, W^7_3]^{\mathrm{T}},\\
  \mathbf{x}^{(2)}&=\mathbf{V^{(2)}}[W^1_6, W^2_1, W^3_3, W^4_5, W^5_7, W^6_2, W^7_4]^{\mathrm{T}}, \\
  \mathbf{x}^{(3)}&=\mathbf{V^{(3)}}[W^1_7, W^2_2, W^3_4, W^4_6, W^5_1, W^6_3, W^7_5]^{\mathrm{T}}.
\end{aligned}
\end{equation*}
Each user is able to get the desired subfiles from the transmissions $\mathbf{x}^{(1)}$, $\mathbf{x}^{(2)}$, and $\mathbf{x}^{(3)}$. Each transmission benefits all the $7$ users.
Consider user $1$. It wants subfiles $W^{1}_5$, $W^1_6$, and $W^1_7$ which are present in $\mathbf{x}^{(1)}$, $\mathbf{x}^{(2)}$, and $\mathbf{x}^{(3)}$, respectively. Let us look at the transmission $\mathbf{x}^{(1)}$. Corresponding to $\mathbf{x}^{(1)}$, the received message  at user $1$ is ${y}^{(1)}_1$, and it is expressed as:
\begin{subequations}
\begin{align}
 {y}^{(1)}_1 & = \mathbf{h}_1^{\mathrm{T}}\mathbf{x}^{(1)}+n_1 \label{eq:ex_consta_1}\\
 &=\mathbf{h}_1^{\mathrm{T}}\mathbf{V^{(1)}}[W^1_5, W^2_7, W^3_2, W^4_4, W^5_6, W^6_1, W^7_3]^{\mathrm{T}} \label{eq:ex_consta_2}\\
 &=\mathbf{h}_1^{\mathrm{T}}\mathbf{v}^{(1)}_1W^{1}_5 + \underbrace{\mathbf{h}_1^{\mathrm{T}}(\mathbf{v}_2^{(1)}W^2_7 + \mathbf{v}_5^{(1)}W^5_6)}_{=0} + \notag  \\
 & \underbrace{\mathbf{h}_1^{\mathrm{T}}( \mathbf{v}_3^{(1)}W^3_2 + \mathbf{v}_4^{(1)}W^4_4 +
 \mathbf{v}_6^{(1)}W^6_1 + \mathbf{v}_7^{(1)}W^7_3)}_\text{user 1 can compute this using $Z_k$} \label{eq:ex_consta_3}
\end{align}
\end{subequations}
where $n_1 \sim \mathcal{C}\mathcal{N}(0,1) $ in \eqref{eq:ex_consta_1} is the additive noise observed at user $1$. Since we consider the high SNR regime, $n_1$ is neglected in the further analysis. The precoding vectors $\mathbf{v}_2^{(1)}, \mathbf{v}_5^{(1)} \in \mathbb{C}^{3 \times 1}$ are designed such that $\mathbf{h}_1^{\mathrm{T}}\mathbf{v}_2^{(1)}=0$ and $\mathbf{h}_1^{\mathrm{T}}\mathbf{v}_5^{(1)}=0$. The precoding vectors and the remaining subfiles, except $W^1_5$, involved in $\mathbf{x}^{(1)}$ are known to user $1$. Thus, $W^1_5$ can be easily decoded from ${y}_1^{(1)}$. Similarly, user $1$ can decode $W^1_6$ and $W^1_7$ from $\mathbf{x}^{(2)}$ and $\mathbf{x}^{(3)}$, respectively. The same procedure applies for other users as well. Thus, the normalized delivery time is obtained as $T_n=3/7$. We have seen that each transmission serves $rt+L=7$ users, and each user is able to retrieve one subfile from one channel use. Thus, the delivery is also one-shot. The optimality of the scheme follows from \eqref{eq:bound}.

\begin{exmp}
	Consider the same setting as in Example \ref{ex:exmp2} with $M=1$.
\end{exmp}
When $M=1$, we have $t=1$, $\gcd(K,t)=1$, and the condition $K=mrt+(m-1)L$ is satisfied with $m=2$. The $(7,7,1,2)$ caching array $\mathbf{C}$ for this case is given in Fig.~\ref{fig:example_t1}. As mentioned earlier, for $t=1$, the columns of $\mathbf{D}$ need not be permuted after filling $\star$'s. Following the $(7,7,1,2)$ caching array $\mathbf{C}$, each subfile is divided into $7$ parts, and the contents stored in each cache $i \in [7]$ are given as $\mathcal{Z}_{C_i}=\{W^n_i, \forall n \in [7]\}$. Then, the contents available to each user are: $Z_1= \{W^n_1,W^n_2, \forall n \in [7]\}$, $Z_2=\{W^n_2,W^n_3, \forall n\in [7]\}$, $Z_3 =\{W^n_3,W^n_4, \forall n \in [7]\}$, $Z_4=\{W^n_4,W^n_5, \forall n \in [7]\}$, $Z_5=\{W^n_5,W^n_6, \forall n \in [7]\}$, $Z_6 =\{W^n_6,W^n_7, \forall n \in [7]\}$, and $Z_7=\{W^n_7,W^n_1, \forall n \in [7]\}$. Consider a distinct demand vector, say $\mathbf{d}=(1,2,3,4,5,6,7)$. There are $7$ transmissions, in total, and each transmission corresponds to an integer $s \in [7]$ in $\mathbf{D}$. In the interest of space, we list only a few transmissions here: 
\begin{equation*}
\begin{aligned}
\mathbf{x}^{(1)}&=\mathbf{V}^{(1)}[W^1_3, W^2_1, W^3_1, W^4_1, W^7_3]^{\mathrm{T}}, \\
\mathbf{x}^{(2)}&=\mathbf{V}^{(2)}[W^1_4, W^2_4, W^3_2, W^4_2, W^5_2]^{\mathrm{T}}, \\
\mathbf{x}^{(3)}&=\mathbf{V}^{(3)}[W^2_5, W^3_5, W^4_3, W^5_3, W^6_3]^{\mathrm{T}}.
\end{aligned}
\end{equation*}
Note that each transmission serves $5$ users. To explain the decoding, consider user $1$ again. User $1$ benefits from $\mathbf{x}^{(1)}$, $\mathbf{x}^{(2)}$, $\mathbf{x}^{(5)}$, $\mathbf{x}^{(6)}$, and $\mathbf{x}^{(7)}$. Let us take the transmission $\mathbf{x}^{(1)}$. The received message at user $1$ corresponding to $\mathbf{x}^{(1)}$ takes the following form (neglecting the additive noise part due to high SNR assumption):
\begin{align*}
 {y}_1^{(1)}& =\mathbf{h}_1^{\mathrm{T}}\mathbf{V}^{(1)}[W^1_3, W^2_1, W^3_1, W^4_1, W^7_3]^{\mathrm{T}}  =\mathbf{h}_1^{\mathrm{T}}\mathbf{v}^{(1)}_1W^1_3 +\\
&  \underbrace{\mathbf{h}_1^{\mathrm{T}}(\mathbf{v}_2^{(1)}W^2_1 + \mathbf{v}_3^{(1)}W^3_1+\mathbf{v}_4^{(1)}W^4_1)}_\text{known to user $1$}    + \underbrace{\mathbf{h}_1^{\mathrm{T}}\mathbf{v}^{(1)}_7W^7_3}_{=0}.
 \end{align*}
User $1$ can thus decode the subfile $W^1_3$. The normalized delivery time achieved in this case is $T_n=1$, which is optimal from \eqref{eq:bound}.

\subsubsection{When $\gcd(K,t)\neq 1$ in Theorem~\ref{thm1}}
\label{subsec:non_unity gcd}
For both the cases in Theorem~\ref{thm1}, $\gcd(K,t)$ needs to be unity. When $\gcd(K,t) \neq 1$,
the optimal delivery time in \eqref{eq:rate} is still achievable for certain specific scenarios. Next, we look at those scenarios. Define $\gamma \triangleq \gcd(K,t,L)$. The optimal normalized delivery time $T^{*}_{n,u}=(K-rt)/(rt+L)$ is achieved with subpacketization level $K/\gamma$ for the following cases: (i) $K=rt+L$ with $\gcd(K/\gamma,t/\gamma)=1$, (ii) $K=mrt+(m-1)L$, $L \geq rt$, $m \in \mathbb{Z}$, $m \geq 2$, and $\gcd(K/\gamma,t/\gamma)=1$. A user grouping strategy is adopted in the above cases to achieve the optimal performance.

Consider case (i) $K=rt+L$ and $\gcd(K/\gamma, t/\gamma)=1$. Construct a $(K, K/\gamma, t/\gamma,r)$ caching array $\mathbf{C}$ and a $(\mathbf{C}, L/\gamma,L)$ delivery array $\mathbf{D}$. Each column of $\mathbf{C}$ and $\mathbf{D}$ can be represented as $k = (i-1)\frac{K}{\gamma}+u$, where $i \in [\gamma]$ and $u \in [{K}/{\gamma}]$. The caching array $\mathbf{C}$ is constructed as:
\begin{equation}
c_{j,k} =\star, \forall j \in \big[\big\langle\frac{(u-1)t}{\gamma}+1\big\rangle_{\frac{K}{\gamma}}, \big\langle \frac{ut}{\gamma}\big\rangle_{\frac{K}{\gamma}}\big],
\label{eq:nonunity_place}
\end{equation}
where each column $\mathbf{c}_k$, $k=(i-1)\frac{K}{\gamma}+u$, contains $t/\gamma$ number of $\star$'s. From \eqref{eq:nonunity_place}, it is evident that the position of $\star$'s in $\mathbf{c}_k$ depends only on $u$. Therefore, for a $u \in [K/\gamma]$, there are $\gamma$ columns with identical $\star$ placement. Condition \textit{B2} is satisfied by \eqref{eq:nonunity_place}. Next, using $\mathbf{C}$, we construct a $(\mathbf{C},L/\gamma,L)$ delivery array $\mathbf{D}$ of size $K/\gamma \times K$ as:
\begin{equation}
\small
d_{j,k} = \star, \forall j \in \bigcup_{l \in [u,\langle u+r-1\rangle_{\frac{K}{\gamma}}]} \big[\big\langle\frac{(l-1)t}{\gamma}+1\big\rangle_{\frac{K}{\gamma}}, \big\langle \frac{lt}{\gamma}\big\rangle_{\frac{K}{\gamma}}\big],
\label{eq:nonunity_useraccess}
\end{equation}
where $k=(i-1)K/\gamma+u$, $i \in [\gamma]$, $u \in [K/\gamma]$. The columns of $\mathbf{D}$ can be split into $\gamma$ groups based on the value of $i$. Each group, denoted by $\mathcal{G}_i$, $i  \in [\gamma]$, contains $K/\gamma$ columns as defined by:
\begin{equation}
\mathcal{G}_i =\{k: k=(i-1)K/\gamma+u, \forall u \in [K/\gamma]\}.
\label{eq:user_group}
\end{equation}
\noindent Then, a column permutation is performed among the columns present in each group $\mathcal{G}_i$ to get the cyclic structure for $\star$'s present in those $K/\gamma$ columns. Thus, the permuted array $\mathbf{D}^{\prime}$ exhibits the following structure
\[
\begin{footnotesize}
\left[
\begin{array}{cccc|cccc|ccc}
\star &  & \cdots & \star & \star &  & \cdots & \star &  & &\\
\star & \star & \cdots & \vdots & \star & \star & \cdots & \vdots & & &\\
\vdots & \star & \cdots &  & \vdots & \star & \cdots &  &  & \cdots & \\
& \vdots & \ddots &  &  & \vdots & \ddots &  &  & &\\
&  & \cdots & \star & & & \cdots & \star &  & &\\
\end{array}
\right].
\end{footnotesize}
\]

The array $\mathbf{D}^{\prime}$ is filled using $L/\gamma$ distinct integers. Consider an integer $s \in [L/\gamma]$, then
\begin{equation}
d_{j,k}^{\prime}=s,\textrm{ } \forall \textrm{ }\langle k \rangle_{\frac{K}{\gamma}}=u \textrm{ and } j = \langle {rt}/{\gamma}+s+u-1\rangle_{\frac{K}{\gamma}},
\label{eq:nonunity_int}
\end{equation}
where  $k=(i-1)K/\gamma+u$, $i \in [\gamma]$, $u \in [K/\gamma]$. Similar to the $\star$'s placement, the integer assignment also depends only on $u$.  It is easy to see that condition \textit{D2} is satisfied by  $\mathbf{D}^{\prime}$. To verify condition $\textit{D3}$, let us take a sub-array $\mathbf{D}^{\prime}_{[1:K/\gamma]}=[\mathbf{d}_1^{\prime},\mathbf{d}_2^{\prime},\ldots,\mathbf{d}_{\frac{K}{\gamma}}^{\prime}]$. The sub-array $\mathbf{D}^{\prime}_{[1:K/\gamma]}$ is of size $K/\gamma \times K/\gamma$, and each of its row and column contains $L/\gamma$ integers.
Then, for an integer $s \in [L/\gamma]$, the sub-array $\mathbf{D}^{\prime(s)}_{[1:K/\gamma]}$ formed by the rows and columns containing $s$ is same as $\mathbf{D}^{\prime}_{[1:K/\gamma]}$. The array $\mathbf{D}^{\prime}$, in fact, appears as a $\gamma-$times duplication of $\mathbf{D}^{\prime}_{[1:K/\gamma]}$ horizontally. Hence, the sub-array $\mathbf{D}^{\prime(s)}$ is same as $\mathbf{D}^{\prime}$, and each row in $\mathbf{D}^{\prime}$ contains $L$ integers. Thus, condition \textit{D3} is also satisfied. Once $\mathbf{D}^{\prime}$ is constructed, the columns are permuted back to obtain $\mathbf{D}$, which is a $(\mathbf{C},L/\gamma,L)$ delivery array. The above constructed $(K,K/\gamma,t/\gamma,r)$ caching array $\mathbf{C}$ and $(\mathbf{C},L/\gamma,L)$ delivery array $\mathbf{D}$ together result in a multi-antenna MACC scheme with normalized delivery time $T_n =\frac{L/\gamma}{K/\gamma}=\frac{ K-rt}{rt+L}$, and subpacketization level $K/\gamma$.

For case (ii) also, an approach similar to case (i) is followed. In this case, we need to construct a $(K,K/\gamma,t/\gamma,r)$ caching array $\mathbf{C}$ and a $(\mathbf{C},(m-1)K/\gamma,L)$ delivery array $\mathbf{D}$. The construction of $\mathbf{C}$ follows \eqref{eq:nonunity_place}. In case (i), we have seen that the array $\mathbf{D}^{\prime}$ appears like a $\gamma-$times duplication of a $K/\gamma \times K/\gamma$ array. Therefore, using the same idea, we construct a $(\mathbf{C},(m-1)K/\gamma,L)$ delivery array $\mathbf{D}$ from a $(\mathbf{C},(m-1)K/\gamma,L/\gamma)$ delivery array $\mathbf{G}$ constructed according to \eqref{eq:s_evena} -- \eqref{eq:s_oddb}. The array $\mathbf{G}$ is replicated $\gamma$-times horizontally to obtain the $(\mathbf{C},(m-1)K/\gamma,L)$ delivery array $\mathbf{D}$.

From the $(K,K/\gamma,t/\gamma,r)$ caching array $\mathbf{C}$ and the $(\mathbf{C},(m-1)K/\gamma,L)$ delivery array $\mathbf{D}$, a multi-antenna MACC scheme can be obtained with $T_n=\frac{(m-1)K/\gamma}{K/\gamma}=\frac{K-rt}{rt+L}$, and subpacketization level $K/\gamma$.

\subsubsection{{A further generalization of Theorem~\ref{thm1}}}
\label{schm_linear}
We show that the multi-antenna scheme in \cite{STS} for dedicated cache networks can be used to obtain a multi-antenna coded caching scheme for MACC networks having parameters $K$, $L$, $r$, $t=KM/N \in [0,\floor{\frac{K}{r}}]$ such that $L \geq rt$ and $\gcd(K,t)=1$. The resulting multi-antenna MACC scheme achieves the normalized delivery time $\frac{K-rt}{rt+L}$ with a subpacketization level requirement in the order of $K$. In this case as well, we first need to construct a $(K,K,t,r)$ caching array $\mathbf{C}$ as in \eqref{eq:star_plac}. Following it, construct another $K \times K$ array $\mathbf{D}$ according to \eqref{eq:user_access}, which represents the side-information available to each user. Each column in $\mathbf{D}$ contains $rt$ number of $\star$'s. Note that array $\mathbf{D}$ is not a delivery array. Then, as done in Theorem~\ref{thm1} and Section~\ref{subsec:non_unity gcd}, permute the columns of $\mathbf{D}$ to obtain a cyclic structure for $\star$'s. Let the permuted version be $\mathbf{D}^{\prime}$. Then using $\mathbf{D}^{\prime}$, the delivery algorithm in \cite{STS} is invoked for transmissions. The delivery algorithm in \cite{STS} requires each subfile to be further split into $rt+L$ parts, and there are, in total, $K(K-rt)$ transmissions. Each transmission is of a mini-subfile size. Thus, we obtain a multi-antenna MACC scheme with $T_n=T^{*}_{n,u}=\frac{K-rt}{rt+L}$ and subpacketization level $K(rt+L)$. This scheme accounts for more instances than the cases considered in Theorem~\ref{thm1}.

\subsection{Construction IV}
\label{subsec:const4}
Construction IV presents a procedure to construct a pair of caching and delivery arrays from any EPDA. From a $(K^{\prime},L^{\prime},F^{\prime},Z^{\prime},S^{\prime})$ EPDA $\mathbf{A}$, we obtain a multi-antenna coded caching scheme for a cyclic wrap-around MACC network having parameters $r$, $K=rK^{\prime}$, $L=rL^{\prime}$, and $\frac{M}{N}=\frac{Z^{\prime}}{rF^{\prime}}$ by constructing a $(K,rF^{\prime}, Z^{\prime},r)$ caching array $\mathbf{C}$ and a $(\mathbf{C},rS^{\prime},L)$ delivery array $\mathbf{D}$. Thus, we have the following lemma.

\begin{lem}
  For any given $(K^{\prime},L^{\prime},F^{\prime},Z^{\prime},S^{\prime})$ EPDA $\mathbf{A}$, there exists a multi-antenna coded caching scheme for a cyclic wrap-around MACC network with each user accessing $r$ caches, $K=rK^{\prime}$ users and caches, $L=rL^{\prime}$ antennas, and $t=\frac{K^{\prime}Z^{\prime}}{F^{\prime}}$. The normalized delivery time achieved by the obtained scheme is $T_n = S^{\prime}/F^{\prime}$, and the subpacketization level of the scheme is $rF^{\prime}$.
  \label{lem:lemma2}
 \end{lem}
\begin{IEEEproof}
	Consider a  $(K^{\prime},L^{\prime},F^{\prime},Z^{\prime},S^{\prime})$ EPDA $\mathbf{A}$. From  $\mathbf{A}$, we construct a $(K=rK^{\prime}, rF^{\prime}, Z^{\prime},r)$ caching array $\mathbf{C}$ and a $(\mathbf{C},S=rS^{\prime},L=rL^{\prime})$ delivery array $\mathbf{D}$ as described below.
	
	To construct the $(K, rF^{\prime}, Z^{\prime},r)$ caching array $\mathbf{C}$, we first construct an intermediate array $\mathbf{C}^{\prime}=[\mathbf{c}^{\prime}_1, \mathbf{c}^{\prime}_2,\ldots, \mathbf{c}^{\prime}_K]$ of size $F^{\prime} \times K$ containing $\star$'s and nulls from $\mathbf{A}$ as follows:  
	\begin{equation}
	\mathbf{c}^{\prime}_{kr} = \mathbf{a}_{k}, \quad \forall k \in [K^{\prime}].
	\label{eq:intarray}
	\end{equation}
	All other columns of $\mathbf{C}^{\prime}$ except those defined in \eqref{eq:intarray} contain only nulls. From $\mathbf{C}^{\prime}$, generate a set of $r-1$ matrices $\mathbf{C}^{\prime}_1, \mathbf{C}^{\prime}_2,\ldots,\mathbf{C}^{\prime}_{r-1}$, each of size $F^{\prime} \times K$, as described in the following: each matrix $\mathbf{C}^{\prime}_i$, $i \in [r-1],$ is obtained by cyclically shifting the columns of $\mathbf{C}^{\prime}$ $i$ times from left to right. Then,  construct  $\mathbf{C}=[\mathbf{C}^{\prime}; \mathbf{C}^{\prime}_1; \mathbf{C}^{\prime}_2;\ldots;\mathbf{C}^{\prime}_{r-1}]$ by vertically concatenating $\mathbf{C}^{\prime}$ and its $r-1$ cyclically shifted versions $\mathbf{C}^{\prime}_1, \mathbf{C}^{\prime}_2,\ldots,\mathbf{C}^{\prime}_{r-1}$. Thus, the array $\mathbf{C}$ is an $rF^{\prime} \times K$ array with each column containing $Z^{\prime}$ stars. Also, from the construction of $\mathbf{C}^{\prime}$ and $\mathbf{C}$, it is easy to see that $\mathbf{C}$ satisfies condition $B2$. Thus, the array $\mathbf{C}$ is a $(K,rF^{\prime},Z^{\prime},r)$ caching array. 
	
	The $(\mathbf{C},rS^{\prime},rL^{\prime})$ delivery array $\mathbf{D}$ of size $rF^{\prime} \times K$ has $rZ^{\prime}$ $\star$'s in each column of $\mathbf{D}$. The array $\mathbf{D}$ can also be partitioned as $\mathbf{D}=[\mathbf{D}^{\prime};\mathbf{D}^{\prime}_1;\mathbf{D}^{\prime}_2;\ldots;\mathbf{D}^{\prime}_{r-1}]$, where $\mathbf{D}^{\prime}$ and $\mathbf{D}^{\prime}_i$, $i \in [r-1]$, are $F^{\prime} \times K$ submatrices of $\mathbf{D}$ formed by the first $F^{\prime}$ rows and rows from $iF^{\prime}+1$ to $(i+1)F^{\prime}$, respectively. Now, consider the submatrix $\mathbf{D}^{\prime}$ which is composed of the first $F^{\prime}$ rows of $\mathbf{D}$.
	The position of $\star$'s in $\mathbf{D}^{\prime}$ is exactly identical to the $\star$'s in an array $\mathbf{A}^{\prime}$ which is obtained by repeating each column of the EPDA $\mathbf{A}$ by $r$ times. Therefore, the vacant cells in $\mathbf{D}^{\prime}$ are filled according to $\mathbf{A}^{\prime}$, and thus we obtain $\mathbf{D}^{\prime}=\mathbf{A}^{\prime}$. 
	 The vacant cells in $\mathbf{D}^{\prime}_i$, $i \in [r-1]$, are filled using the set of integers $[iS^{\prime}+1,(i+1)S^{\prime}]$ by cyclically shifting each integer in $\mathbf{D}^{\prime}$ by $i$ times and then adding $iS^{\prime}$ to it. Thus, we obtained the array $\mathbf{D}=[\mathbf{D}^{\prime};\mathbf{D}^{\prime}_1;\mathbf{D}^{\prime}_2;\ldots;\mathbf{D}^{\prime}_{r-1}]$ with $rZ^{\prime}$ $\star's$ in each column and $rS^{\prime}$ integers, in total. It remains to verify whether the array $\mathbf{D}$ satisfies the conditions $D2$ and $D3$. 
	Note that it is enough to show that the submatrix $\mathbf{D}^{\prime}$ satisfies the conditions $D2$ and $D3$. Since the integers are filled according to the EPDA $\mathbf{A}$, it is guaranteed that no appears integer more than once in any column of $\mathbf{D}^{\prime}$ and 
	the set of integers present in each of the $r$ submatrices defined above are different. Thus, condition $D2$ is satisfied. To verify condition $D3$, consider an integer $s \in [S^{\prime}]$ and the corresponding sub-array $\mathbf{D}^{\prime(s)}$. No row in $\mathbf{D}^{\prime(s)}$ contains more than $rL^{\prime}$ integers as $\mathbf{D}^{\prime (s)} =\mathbf{A}^{\prime(s)}$ (recall that $\mathbf{A}^{\prime}$ is constructed by replicating each column of the EPDA $\mathbf{A}$ by $r$ times). Thus, condition $D3$ is satisfied and $\mathbf{D}$ is a $(\mathbf{C},rS^{\prime},rL^{\prime})$ delivery array. 
	
	With the above constructed $(rK^{\prime},rF^{\prime},Z^{\prime},r)$ caching array $\mathbf{C}$ and $(\mathbf{C}, rS^{\prime},rL^{\prime})$ delivery array $\mathbf{D}$, we obtain a multi-antenna coded caching scheme for a cyclic wrap-around MACC network with $rK^{\prime}$ users and caches, and each cache has a normalized size $\frac{M}{N}=\frac{Z^{\prime}}{rF^{\prime}}$  using Lemma \ref{ref:lem}. The obtained multi-antenna coded caching scheme achieves the normalized delivery time $T_n = \frac{rS^{\prime}}{rF^{\prime}}=\frac{S^{\prime}}{F^{\prime}}$ with a subpacketization level $rF^{\prime}$.
	
	If the EPDA that we begin with is $g-$regular, then each integer in the obtained delivery array appears $rg$ times. Therefore, if the EPDA is $(t+L^{\prime})-$regular (or in other words, $(\frac{K^{\prime}Z^{\prime}}{F^{\prime}}+L^{\prime})-$regular), then each integer in $\mathbf{D}$ appears $rt+L$ times. Consequently, each transmission in the delivery scheme  benefits $rt+L$ users.
\end{IEEEproof}
Using the EPDAs obtained from Construction III in \cite{NPR} in Lemma \ref{lem:lemma2}, we have the following theorem.

\begin{thm}
	From a $(t+L^{\prime})-$regular $(K^{\prime},L^{\prime}, (t+L^{\prime})\binom{K^{\prime}}{t+L^{\prime}},t\binom{K-1}{t+L^{\prime}-1},(K^{\prime}-t)\binom{K^{\prime}}{t+L^{\prime}})$ EPDA $\mathbf{A}$, a multi-antenna coded caching scheme can be obtained for a cyclic wrap-around MACC network with each user accessing $r$ caches, $K = rK^{\prime}$ users, and $L=rL^{\prime}$ antennas. The normalized delivery time achieved by the scheme is 
	\begin{equation}
	   T_n = \frac{K-rt}{rt+L}.
	\end{equation}
	 \label{thm3}
\end{thm}
\begin{IEEEproof}
 As described in the proof of Lemma \ref{lem:lemma2}, we construct a $(rK^{\prime},r(t+L^{\prime})\binom{K^{\prime}}{t+L^{\prime}},t\binom{K-1}{t+L^{\prime}-1},r)$ caching array $\mathbf{C}$ and a $(\mathbf{C},r(K^{\prime}-t)\binom{K^{\prime}}{t+L^{\prime}},rL^{\prime})$ delivery array $\mathbf{D}$ from the $(t+L^{\prime})-$regular $(K^{\prime},L^{\prime}, (t+L^{\prime})\binom{K^{\prime}}{t+L^{\prime}},t\binom{K-1}{t+L^{\prime}-1},(K^{\prime}-t)\binom{K^{\prime}}{t+L^{\prime}})$ EPDA $\mathbf{A}$. The above pair of arrays together describe a multi-antenna coded caching scheme for an MACC network with $K = rK^{\prime}$ users and $L=rL^{\prime}$ antennas. The scheme achieves a normalized delivery time $T_n = \frac{K^{\prime}-t}{t+L^{\prime}}=\frac{K-rt}{rt+L}$, which is optimal by \eqref{eq:bound}. The subpacketization level of the scheme is $(rt+L)\binom{K^{\prime}}{t+L^{\prime}}$.
\end{IEEEproof}

It is important to note that the scheme in Theorem \ref{thm3} completely characterizes the optimal normalized delivery time versus memory tradeoff (under uncoded placement and one-shot delivery) of a multi-antenna MACC network satisfying the conditions $r \mid K$ and $r \mid L$.

We now present an example to illustrate the scheme in Theorem \ref{thm3}.

\begin{exmp}
 Consider a cyclic wrap-around MACC network having $K=10$ users and caches, and each cache has a normalized size $M/N = 1/5$. Each user has access to $r=2$ caches in a consecutive and cyclic wrap-around manner. The server has a library of $N$ files and has $L=4$ transmit antennas. 
 \end{exmp}

\begin{figure*}[t!]
	\centering
	\includegraphics[width=0.58\textwidth]{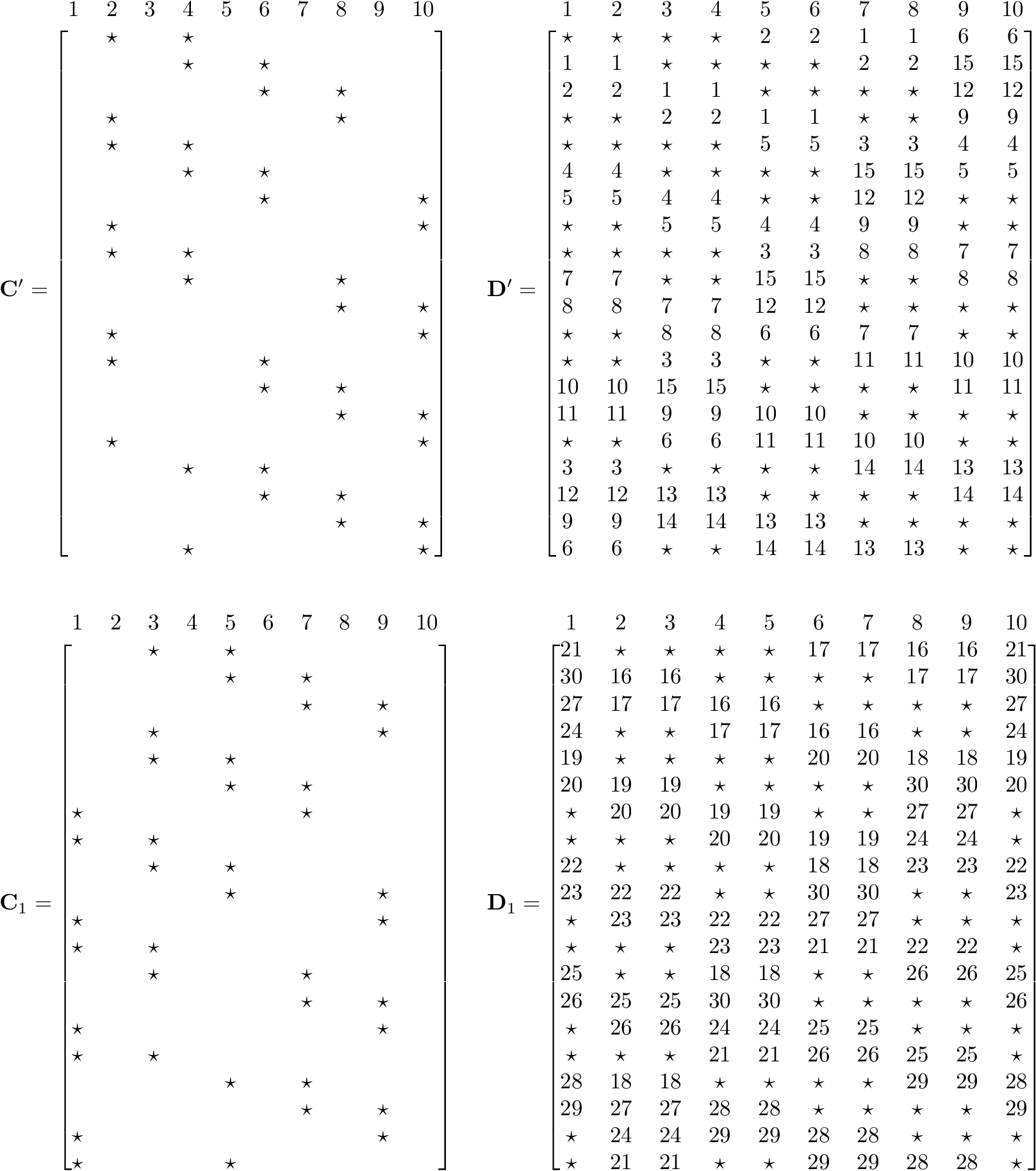}
	\caption{Caching and delivery arrays for $K=10$, $r=2$, $L=4$, $t=2$.}
	\label{fig:example_thm3}
\end{figure*}

To construct caching and delivery arrays for the above case, we start with a $(5,2,20,8,15)$ EPDA given in \eqref{eq:epdaexmp}. From $\mathbf{A}$, we construct a $(10,40,8,2)$ caching array $\mathbf{C}$, where $\mathbf{C}=[\mathbf{C}^{\prime};\mathbf{C}_1]$. The matrices $\mathbf{C}^{\prime}$ and $\mathbf{C}_1$ are given in Fig. \ref{fig:example_thm3}.

Each file $W^n, n \in [N]$, is split into 40 subfiles, and each cache stores $8$ subfiles of every file in the library. Thus, $M=N \times 8/40=N/5$. The $(\mathbf{C},30,4)$ delivery array $\mathbf{D}$ is constructed from the matrices $\mathbf{D}^{\prime}$ and $\mathbf{D}_1$ illustrated in Fig. \ref{fig:example_thm3} as $\mathbf{D}=[\mathbf{D}^{\prime}; \mathbf{D}_1]$. Each user $k \in [10]$ has access to 16 subfiles of every file. Thus, the effective cache size seen by each user is $2N/5$. When the demand vector $\mathbf{d}$ is known, the server sends a transmission corresponding to every integer in $\mathbf{D}$. Note that each integer appears $rt+L=8$ times in $\mathbf{D}$. Therefore, each transmission from the server benefits $8$ users, and the normalized delivery time obtained is $T_n  = 3/4$, which is optimal from \eqref{eq:bound}.

\begin{equation}
\begin{footnotesize}
\mathbf{A}=\left[\begin{array}{ccccccccc}
\star & \star & 2 & 1 & 6 \\
1 & \star & \star & 2 & 15 \\
2 & 1 & \star & \star & 12 \\
\star & 2 & 1 & \star & 9 \\
\star & \star & 5 & 3 & 4 \\
4 & \star & \star & 15 & 5\\
5 & 4 & \star & 12 & \star \\
\star & 5 & 4 & 9 &  \star \\
\star & \star & 3 & 8 & 7 \\
7 & \star & 15 & \star & 8 \\
8 & 7 & 12 & \star & \star \\
\star & 8 & 6 & 7 & \star \\
\star & 3 & \star & 11 & 10 \\
10 & 15 & \star & \star & 11 \\
11 & 9 & 10 & \star & \star \\
\star & 6 & 11 & 10 & \star \\
3 & \star & \star & 14 & 13 \\
12 & 13 & \star & \star & 14 \\
9 & 14 & 13 & \star & \star \\
6 & \star & 14 & 13 & \star  
\end{array}\right]
\label{eq:epdaexmp}
\end{footnotesize}
\end{equation}

As a direct consequence of Construction IV, we have the following remark (Remark \ref{const4_conseq}).

\begin{rem}
	If a $(\frac{K^{\prime}Z^{\prime}}{F^{\prime}}+L^{\prime})-$regular EPDA  is used in Lemma \ref{lem:lemma2}, we can obtain a multi-antenna coded caching scheme for a cyclic wrap-around MACC network with parameters $r$, $K =rK^{\prime}$ users, and $L = rL^{\prime}+\ell$ transmit antennas, where $\ell$ is an integer such that $\ell < r$. Then, each transmission in the delivery phase benefits $rt+L^{\prime}r$ users and $T_n = \frac{K-rt}{rt+L-\ell}$.
	\label{const4_conseq}
\end{rem}


\begin{rem}
	
The delivery arrays proposed in this work are, in fact, EPDAs with $\star$'s defined by another array called the caching array. Specifically, in the delivery array, the number of $\star$'s in every row is a multiple of $r$, and the $\star$'s appear in blocks of length $r$. This is a consequence of the users' accessing pattern of caches, i.e., a subfile cached in one cache will be accessed by $r$ neighbouring users. Thus, being an EPDA is not sufficient to be a delivery array. However, certain EPDAs are delivery arrays as well. Interestingly, the two classes of EPDA constructions in \cite{NPR} are modified into delivery arrays in Constructions II and III. Later, in Construction IV, we present a procedure to obtain a pair of caching and delivery arrays from any EPDA. When $r=1$, the network model reduces to a single-access multi-antenna setting, and all the four proposed multi-antenna MACC schemes reduce to multi-antenna schemes for dedicated cache networks in \cite{NPR} that achieve optimal performance under uncoded placement and one-shot delivery. It is important to note that the scheme resulting from Construction I recovers the multi-antenna scheme in \cite{NPR} as a special case, but Construction I is not based on any existing EPDA construction.
	
\end{rem}
	The multi-antenna MACC schemes resulting from Constructions II, III, and IV carry significance because of their optimal performance under uncoded placement and one-shot delivery. In the single-antenna setting, there exists only a scheme that achieves optimal performance under uncoded placement. The above single-antenna multi-access scheme exists when the condition $r=(K-1)/t$ is satisfied. In fact, the multi-antenna MACC scheme resulting from Construction II subsumes this condition as a special case. Further analyses on the performance of the proposed schemes are given in the following section.

\section{Performance Comparison and Analyses}
{As mentioned earlier, this is the first work considering multi-antenna setting in cyclic wrap-around MACC networks. All our proposed multi-antenna MACC schemes are obtained from caching and delivery arrays that are appropriately designed for the respective scenarios. Hence, all our schemes follow an uncoded placement and one-shot delivery. We have already shown that the schemes proposed in Theorem~\ref{thm1} and Theorem~\ref{thm3} achieve optimal performance under uncoded placement and one-shot delivery. Note that schemes in Theorem~\ref{thm1} achieve optimal performance with a subpacketization level $K$. For single-antenna cyclic wrap-around MACC networks, there are only two schemes \cite{NaR,SPE} that achieve optimal performance  under some specific constraints (the scheme in \cite{NaR} is for lower memory regime and consider $N \leq K$ setting. Note that the scheme in \cite{NaR} achieves optimal performance using coded placement; whereas, the scheme in \cite{SPE} achieves optimal performance under uncoded placement when $r = {(K-1)}/{t}$). In fact, the scheme for case $(a)$
 in Theorem~\ref{thm1} recovers the scheme in \cite{SPE} when $L=1$. It is worth noting that there exists no general scheme that achieves optimal performance in single-antenna setting even under the constraint of uncoded placement. Therefore, constructions II, III, and IV are significant as they result in multi-antenna MACC schemes that are optimal at several instances.

 \subsubsection{Comparison with the scheme in \cite{CWLZC} when $L=1$}
{The general scheme in Theorem~\ref{thm_general} achieves the normalized delivery time $T_n=\frac{K-rt}{t+L}$ for any $K \geq r(t+L)$, with a subpacketization level  $K\binom{K-(r-1)(t+L)-1}{t+L-1}$. We have already shown that the subpacketization level can be further reduced if $\gcd(K,t,L) \neq 1$. From the normalized delivery time expression, it is evident that the scheme provides full local caching gain and full multiplexing gain but does not achieve the full global coded caching gain. However, when $L=1$, our scheme reduces to a single-antenna cyclic wrap-around MACC scheme having a normalized delivery time $\frac{K-rt}{t+1}$  and a subpacketization level $K\binom{K-(r-1)(t+1)-1}{t}$ (Note that in single-antenna setting, the performance measure is called as rate and is equivalent to the normalized delivery time in the multi-antenna setting). We compare the performance of our single-antenna MACC scheme resulting from Theorem~\ref{thm_general} with that of the scheme in \cite{CWLZC}. The scheme in \cite{CWLZC} also achieves the normalized delivery time $\frac{K-rt}{t+1}$, but requires a subpacketization level $K\binom{K-t(r-1)}{t}$ which is higher than the subpacketization level required in our scheme ($K\binom{K-(r-1)(t+1)-1}{t}$). Figure~\ref{fig:comp_L1} presents a comparison on the subpacketization level requirements of our scheme (obtained from Theorem~\ref{thm_general}) and the scheme in \cite{CWLZC} by considering an example with $K=25$, $N=25$, and $r=3$. Note that our scheme offers a significant gain in the subpacketization level as $M$ increases.}
 \begin{figure}[t!]
 	\begin{center}
 		\captionsetup{justification=centering}
 		\includegraphics[width=0.76\columnwidth]{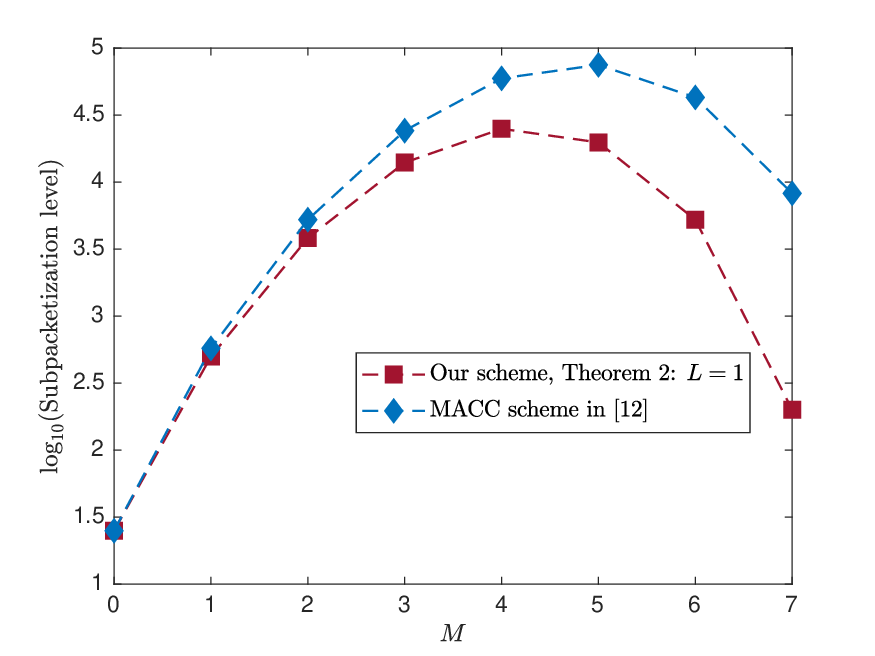}
 		\caption{Cyclic wrap-around MACC schemes: \\$K=25$, $N=25$, $r=3$, $L=1$}
 		\label{fig:comp_L1}
 	\end{center}
 \end{figure}
 
 \subsubsection{Comparison with the scheme in \cite{CBWC}}
 Recently, a multi-antenna coded caching scheme has been proposed for cyclic wrap-around MACC networks in \cite{CBWC}. The scheme in \cite{CBWC} is derived from a PDA, and the authors show that corresponding to every $(t+1)-$regular $(K^{\prime},F^{\prime},Z^{\prime},S^{\prime})$ PDA, there exists a multi-antenna coded caching scheme for the cyclic wrap-around MACC network having $K = (r-1)K^{\prime}+t$ users, $L$ transmit antennas, and $t=\frac{K^{\prime}Z^{\prime}}{F^{\prime}}$. Note that the scheme in \cite{CBWC} works only when $r>L$ and $\frac{K-t}{r-1} \in \mathbb{N}$, and the normalized delivery time achieved is $T_n=\frac{K-rt}{Lt+L}$ with a subpacketization level $KF^{\prime}L/\gcd(L,r-1)$. Since $r >L$, only the schemes in Theorem \ref{thm_general} and Theorem \ref{thm1} (case (a)) need to be compared with the scheme in \cite{CBWC}. The scheme in \cite{CBWC} achieves a higher DoF $Lt+L$ compared to the $t+L$ DoF achieved by the scheme in Theorem \ref{thm_general} (Construction I). Therefore, the scheme in \cite{CBWC} has a lower $T_n$ compared to the scheme in Theorem \ref{thm_general}.

 	The scheme in case (a) in Theorem \ref{thm1} (resulting from Construction II: $K=rt+L$, $\gcd(K,t)=1$) and the scheme in \cite{CBWC} are together applicable when $L=r-1$, $\gcd(K,t)=1$, and $K^{\prime}=t+1$. In all those scenarios, the scheme from Construction II achieves the optimal performance $T_n=\frac{K-rt}{rt+L}$ with a subpacketization level $K$. In contrast, the scheme  in \cite{CBWC} achieves $T_n=\frac{K-rt}{Lt+L}$ with a subpacketization level $K(t+1)$. Thus, our scheme from Construction II outperforms the scheme in \cite{CBWC} both in terms of $T_n$ and subpacketization level. For example, consider an MACC network with $K=19$, $r=5$, $L=4$, and $t=3$. The scheme in \cite{CBWC} has $T_n = 1/4$ with a subpacketization level 76. On the other hand, the scheme from Construction II achieves $T_n = 4/19$ with a subpacketization level $19$.
 	
 
 Next, using an example, we  illustrate that our scheme in Theorem \ref{thm_general} provides an advantage in the subpacketization level over the scheme in \cite{CBWC}. Consider an example with $N=20$, $K=20$, $r=3$, $t=4$, and $L=2$. Since the above example satisfies the conditions $r > L$ and $K \geq r(t+L) $, the schemes in \cite{CBWC} and Theorem \ref{thm_general} can be used. The scheme in \cite{CBWC} achieves the normalized delivery time $T_n = 4/5$ with a subpacketization level $1400$ using a $5-$regular $(8,70,35,35)$ PDA. Whereas, the scheme in Theorem \ref{thm_general} achieves $T_n = 4/3 $ with a subpacketization level $410$. Since $\gcd(K,t,L)=2$, we can further reduce the subpacketization level to $30$ by employing the approach described in Section \ref{subsubsec:subredn} without affecting $T_n$. 
 

 \begin{figure}[t]
 	\begin{center}
 		\captionsetup{justification=centering}
 		\includegraphics[width=0.76\columnwidth]{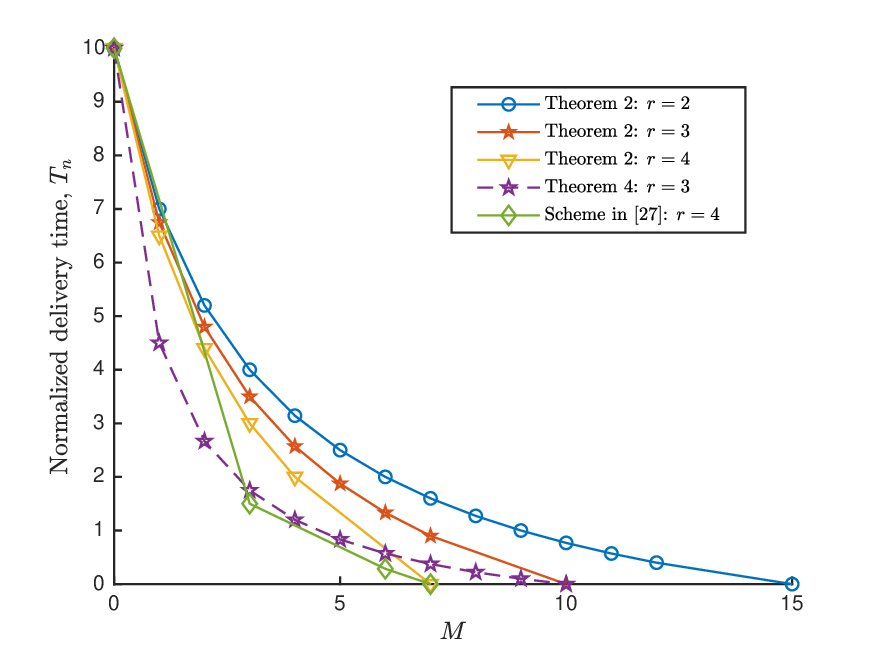}
 		\caption{Cyclic wrap-around MACC schemes: \\$K=30$, $N=30$, $L=3$}
 		\label{fig:comp_rate}
 	\end{center}
 \end{figure}

\begin{table*}[t!]
	\centering
	\caption{Known multi-antenna schemes for the cyclic wrap-around MACC network with parameters $K$, $r$, $t$, and $L$}
	\begin{tabular}{|c | c | c | c|} 
		\hline
		Schemes & Constraints & Normalized delivery time, $T_n$ & Subpacketization level \\ [0.5ex] 
		\hline\hline
		\rule{0pt}{2.6ex}
		\multirow{2}{*}{Construction I} & $K \geq r(t+L)$ & $\frac{K-rt}{t+L}$ & $K\binom{K-(r-1)(t+L)-1}{t+L-1}$ \\ 
		\cline{2-4}
		\rule{0pt}{3.6ex}
		& \makecell{$K \geq r(t+L), \gcd(K,t)=\gamma > 1$, \\ $K^{\prime}=\frac{K}{\gamma}$, $t^{\prime}=\frac{t}{\gamma}$, $L^{\prime}=\frac{L}{\gamma}$} & $\frac{K-rt}{t+L}$ & ${K^{\prime}\binom{K^{\prime}-(r-1)(t^{\prime}+L^{\prime})-1}{t^{\prime}+L^{\prime}-1}}$\\ 
		\hline
		Construction II & $K=rt+L,$ $\gcd(K,t)=1$ & $\frac{K-rt}{rt+L}$ & $K$ \\
		\hline 	\rule{0pt}{3.6ex}
		Construction III & \makecell{$K=mrt+(m-1)L,$ $L \geq rt,$ \\ $\gcd(K,t)=1$, $m \in \mathbb{N} \backslash \{1\}$} & $\frac{K-rt}{rt+L}$ & $K$  \\
		\hline \rule{0pt}{2.6ex}
		\makecell{Scheme in Section \ref{schm_linear} } & $L \geq rt$, $\gcd(K,t)=1$ & $\frac{K-rt}{rt+L}$ & $K(rt+L)$ \\
		\hline \rule{0pt}{3.6ex}
		\makecell{Construction IV } & $r \mid K$, $r \mid L$ & $\frac{K-rt}{rt+L}$ & \makecell{$(rt+{L})\binom{{K}/{r}}{t+{L}/{r}}$} \\ 
		\hline \rule{0pt}{3.6ex}
		\makecell{Scheme in Remark \ref{const4_conseq} } &  \makecell{$r \mid K$, $L =rL^{\prime}+\ell$, \\ $L^{\prime},\ell \in \mathbb{N}$, $\ell < r$} & $\frac{K-rt}{rt+L-\ell}$ & \makecell{$r(t+L^{\prime})\binom{K/r}{t+L^{\prime}}$  }   \\
		\hline \rule{0pt}{3.6ex}
		Scheme in \cite{CBWC} & \makecell{${K = (r-1)K^{\prime}+t}$, $r >L$,\\ $K^{\prime} \in \mathbb{N}$ }  & $\frac{K-rt}{Lt+L}$ &  \makecell{$\frac{KL\binom{K^{\prime}}{t}}{\gcd(L,r-1)}$}\\ [1ex] 
		\hline
	\end{tabular}
	\label{table:summary}
\end{table*}

 \subsubsection{Normalized delivery time versus memory tradeoff of the scheme in Theorem \ref{thm_general}}
 When $r=1$, the network model reduces to the multi-antenna dedicated cache network, and in that case, our scheme in Theorem~\ref{thm_general} reduces to the multi-antenna scheme in \cite{NPR}. In Fig.~\ref{fig:comp_rate}, we present the normalized delivery time versus memory tradeoff for a network with $K=30$, $N=30$, $L=3$, and $r$ takes values from two to four. When the number of caches accessible to each user $r$ increases, the normalized delivery time $T_n$ decreases as more contents are locally available to the users from the caches. Note that the scheme in Theorem~\ref{thm_general} requires $K \geq r(t+L)$ which implies $t \leq \floor{\frac{K}{r}-L}$. When $M=\frac{N}{r}$, users get their demanded files directly from their accessible caches, hence $T_n = 0$. When $M=0$, the normalized delivery time required in all the cases is $T_n = {K}/{L}=10$.
 When $r=3$ and $M=9$, the condition $K = rt+L$ is satisfied. Hence, we can achieve the optimal normalized delivery time $T_n=1/10$ using the scheme in case (a) in  Theorem~\ref{thm1}. When $r=3$, the scheme in Theorem $4$ can be applied as $r \mid K$ and $r \mid L$. In that case, the  scheme in Theorem $4$ completely characterizes the optimal normalized delivery times versus memory tradeoff, under uncoded placement and one-shot delivery. Note that the scheme in Theorem $4$ achieves  full local caching gain, full coded caching gain, and full multiplexing gain (optimal performance under uncoded placement and one-shot delivery), whereas, the scheme in Theorem \ref{thm_general} achieves only full local caching gain and full multiplexing gain. When $r=4$ and $\frac{K-t}{r-1}\in \mathbb{N}$, the scheme in \cite{CBWC} is valid. From Fig. \ref{fig:comp_rate}, it is evident that the scheme in \cite{CBWC} achieves a better performance compared to the scheme in 
Theorem~\ref{thm_general} for $2 \leq M \leq 7$.
 
For further comparison, we list all the known multi-antenna schemes for the cyclic wrap-around MACC networks with their performances and constraints in Table \ref{table:summary}. The subpacketization levels mentioned  for Construction IV and Scheme in Remark \ref{const4_conseq} correspond to the subpacketization levels obtained by using EPDAs from Construction III in \cite{NPR}. For the scheme in \cite{CBWC}, we obtained the subpacketization level by using the $(K^{\prime},\binom{K^{\prime}}{t},\binom{K^{\prime}-1}{t-1},\binom{K^{\prime}}{t+1})$ PDA.

\section{Conclusion and Discussion}
We studied the cyclic wrap-around MACC networks with multiple transmit antennas and proposed four multi-antenna MACC schemes for different parameter settings. The schemes are derived from a pair of arrays called the caching array and the delivery array. The first construction resulted in a more general scheme, and the other schemes resulting from the remaining three constructions have a limited applicability compared to the first one. However, the latter three schemes achieved optimal performance under uncoded placement and one-shot delivery. The performance of the proposed schemes is analysed only in the high SNR regime, considering the finite SNR regime is another direction to work on. Also, characterizing the optimal normalized delivery time versus memory tradeoff of the cyclic wrap-around MACC network is another problem to be explored.

\end{document}